\documentstyle[12pt,epsf]{article}
\title{
  {\vspace{-3cm} \normalsize \hfill
    \parbox{38mm}{CERN-TH 98-380\\
                  HD-THEP-98-60 \\
                  FTUAM-98-26 \\}  }\\[25mm]
  Cooling, Physical Scales and Topology
}

\author{
Margarita Garc\'{\i}a P\'erez$^{1,2}$,\\
Owe Philipsen$^{2,3}$\\
and Ion-Olimpiu Stamatescu$^{2,4}$\\
[5mm]
{\small $^1$ Departamento de F\'{\i}sica Te\'orica, 
Universidad Aut\'onoma de Madrid,}\\
{\small Canto Blanco, 28049 Madrid, Spain}\\
{\small $^2$Institut f\"ur Theoretische Physik, Philosophenweg 16,}\\
{\small D-69120, Heidelberg, Germany}\\
{\small $^3$Theory Division, CERN, CH-1211, Geneva 23, Switzerland}\\
{\small $^4$FESt, Schmeilweg 5, D-69118, Heidelberg, Germany}
}
\date{}
\newcommand{\be}{\begin{equation}}
\newcommand{\ee}{\end{equation}}
\newcommand{\bea}{\begin{eqnarray}}
\newcommand{\eea}{\end{eqnarray}}

\newcommand{\Tr}{{\rm Tr}}

\def\NPB{{Nucl.\ Phys.} B}
\def\PLB{{Phys.\ Lett.} B}

\def\PRD{{Phys.\ Rev.} D}

\newcommand{\basispl}{
   \put(-1.,-.5){\line(1,0){2}}
   \put(-1.,.5){\line(1,0){2}}
   \put(-1.,-.5){\line(0,1){1}}
   \put(1.,-.5){\line(0,1){1}}
                         }

\newcommand{\basisal}{
   \put(.0,.5){\vector(1,0){0}}
   \put(.0,-.5){\vector(-1,0){0}}
   \put(-1,.0){\vector(0,1){0}}
   \put(1,.0){\vector(0,-1){0}}
                         }

\newcommand{\twooneplaq}{\setlength{\unitlength}{.5cm}
   \raisebox{-.2cm}{
   \begin{picture}(2.2,1.2)(-1.0,-.6)
   \basispl\basisal
   \put(-1.6,-0.9){$x$}
   \put(1.2,-0.9){$x\!+\!m\mu$}
   \put(-2.3,0.9){$x\!+\!n\nu$}
   \put(1,-.5){\circle*{.2}}
   \put(-1,-.5){\circle*{.2}}
   \put(-1.,.5){\circle*{.2}}
   \end{picture}}}

\begin{document}
\maketitle

\begin{abstract}
We develop a cooling method controlled by a physical {\it cooling radius}
that defines a scale below which fluctuations are smoothed out while leaving
physics unchanged at all larger scales. 
We apply this method to study topological properties of lattice
gauge theories, in particular the behaviour of
instantons, dislocations and instanton--anti-instanton pairs. 
Monte Carlo results for the
$SU(2)$ topology are presented.
We find that the method provides a means to prevent instanton--anti-instanton
annihilation under cooling. While the instanton sizes are largely
independent from the smoothing scale, the density and pair separations
are determined by the particular choice made for this quantity. 
We discuss the questions this raises for the ``physicality" of these
concepts.
\vspace*{2cm}
\flushleft{ CERN-TH 98-380\\ December 1998}
\end{abstract}
\newpage

\section{Introduction}

Beyond providing quantitative estimates for various 
physical parameters like masses and decay constants,
Monte Carlo simulations of lattice gauge
theories are increasingly involved in obtaining a
physical picture of the structure of quantum field theories, such as
the quantum vacuum and its topological properties. 
Such analyses are hampered by the ultraviolet fluctuations of the
fields at scales comparable with the lattice spacing. 
A typical example of these problems is
the scaling behaviour of the SU($N=2,3$)
topological susceptibility which, for the commonly used
lattice actions, turns out to be spoiled by the presence 
of topological fluctuations at the scale of the cut-off (dislocations).

The need to improve the continuum approach of physical quantities 
has motivated the development of techniques to smooth out the UV, 
cut-off-dependent, structure of the configurations before measuring on them 
the quantities of interest, be these physical parameters like masses or 
features of the quantum vacuum.
Apart from the improvement of lattice actions
there are essentially two approaches that have been taken to realize
this program:\par 
(i) preparing operators that are insensitive to UV fluctuations, and\par
(ii) smoothing the fields themselves. 

For example, by smearing the spatial link variables \cite{alb} 
on fixed time slices one 
constructs spatially extended operators  with increased overlap onto
the physical states, and thereby also achieves (i). This procedure
preserves the transfer matrix  and hence the  physical content. 
It has been successfully applied to 
potential and spectrum measurements. 

Smearing of the link variables
can also be performed in an isotropic way, i.e. 
involving also the links in the time direction. 
When this is done on all links 
of the lattice before the measurements, it amounts to (ii). 
In this case locality
and hence the transfer matrix interpretation are 
lost at the level of the original lattice.
Isotropic smearing has been found 
to be useful to minimize UV renormalization
effects of the charge operator in  
measurements of the topological susceptibility \cite{digiat}
and other properties of the gauge theory vacuum \cite{boulder}.
However, it is not always clear if and to which limit the procedure 
converges \cite{phil}.

A special case of the second approach 
is cooling \cite{tep0}. It consists of an 
iterative, local minimization of the action, which 
proceeds sweep-wise
and converges to configurations fulfilling the classical
(Euclidean) equations of motion. In the case of continuum gauge theories, 
these are configurations characterized by the topological
sector $Q$ obeying
$S = |Q|S_1$, where $S_1=8 \pi^2 /g^2$ is the action of one instanton.
Hence, up to scaling violations associated with the lattice discretization and 
assuming that topology preserves its meaning away from the semi-classical
approximation, cooling should reveal the 
topological sector to which the original configuration belongs.

Different smoothing techniques such as smearing \cite{boulder}, 
cooling \cite{schi}-\cite{alvaro} and 
inverse blocking \cite{boulder0,mpr} 
have also been applied in investigating whether
gauge theory configurations can be described as instanton ensembles.
Although results obtained for the topological susceptibility
seem to be  procedure-independent
(i.e. they agree within 10$\%$), the situation is far less satisfactory 
for the determination of the instanton size distribution
or the instanton density, which is
mostly due to ambiguities in extracting the instanton content from the
Monte Carlo configurations (for recent reviews, see \cite{Pierre,Neg}).

Cooling is known to proceed as a diffusion process in
the sense that the length scale up to which it smoothes the configurations
grows with the number of iterations \cite{tepe1,pisa1}.
There are two problematic features that
are common to standard cooling procedures:
\begin{enumerate}
\item
In general the physical spectrum cannot be extracted from correlation
distances shorter than the ``cooling radius" to which smoothing has advanced.
For example, the string tension extracted from a given pair of time slices
rapidly diminishes as cooling proceeds (for discussions of this point,
see below and \cite{gon95,dig89}).
\item
Instanton--anti-instanton pairs, unlike superpositions of
only instantons (I) or only anti-instantons (A), 
do not represent saddle points of the action. They 
belong to the topologically trivial sector and can have
any action between $2S_1$ and 0. 
Consequently they are indistinguishable 
from trivial topological fluctuations and are removed by the 
cooling process.
On the other hand, since in many instanton-based models
they are conjectured to be relevant for important physical effects
(for a review see \cite{Shuryak}) one would like to investigate if and
how they are represented in the configurations.
\end{enumerate}

The question then is whether it is possible to apply a well
defined amount of cooling, such that the configurations do not lose their
physical properties at the length scales one is studying,
but have become smooth enough to allow their identification.
With the present cooling techniques, a well defined criterion to achieve this
is lacking. The amount of cooling employed has to be fixed by 
subjectively judging when the configurations appear smooth
enough to extract topological 
quantities\footnote{ We stress that this difficulty does not affect 
the topological susceptibility, measured on cooled configurations, since I-A 
annihilation does not change the topological sector $Q$.}.
In this paper, we introduce 
a systematic way of determining
the cooling radius 
and relating it directly to a physical (continuum) length 
scale. This allows a
smoothing of the fields up to a predefined scale, leaving
all properties on larger scales untouched, and hence avoids the
necessity of 
monitoring the cooling process, with all uncertainties involved in it.

The procedure we shall introduce and discuss here is to let cooling saturate
according to a {\it local} criterion. Following a suggestion by
Niedermayer \cite{nied}, the local updating of links in the 
direction of minimization of the local action is only performed if the 
equations of motion are locally violated by more than
a threshold $\delta$ (see section 2). 
As a result, cooling will only proceed where the links are still far from
the local minimum, and stop when the structure has everywhere approached the 
classical solution to a degree related to $\delta$.  
As we will see, this allows to preserve the string tension above a 
distance given by the cooling radius and provides a natural 
lattice implementation 
of the valley method constraint to prevent 
instanton--anti-instanton annihilation (see e.g. \cite{Valley}).

In section 2 we introduce our cooling algorithm.
The problem of relating the cooling radius to a continuum length scale 
will be discussed in section 3.
In section 4 we test the effect of the algorithm on small
instantons and I-A pairs, whereas section 5 contains results
for the topological susceptibility and the properties of 
the instanton ensemble.
Finally, section 6 is reserved for conclusions and outlook.

 The numerical results presented in this work have been obtained on the lattices
indicated in Table \ref{t.sim}. 
\begin{table}[ht]
\begin{center}
\label{t.sim}
\begin{tabular}{|c|c|c|c|c|c|c|}
\hline
 
Lattice & b.c.& $\beta$& $a$ (fm)&Confs.& Sep. sweeps&Therm. sweeps  \\ \hline
 
 $12^3\times 36$&Periodic& 2.4  &  0.12& 800&100&20000 \\ \hline
 
 $24^4$ &Twist in time &2.6& 0.06&350&200&20000  \\ \hline
 
\end{tabular}
 
\caption[]{{\it Simulation parameters}}
\end{center}
\end{table}
The $12^3 \times 36$ lattice has periodic 
boundary conditions in all four directions, while the $24^4$ lattice 
has twisted boundary conditions
in the time direction, with twist $\vec{k}=(1,1,1)$. The
large ``time'' extension  and the twisted boundary conditions have been 
adopted in order to suppress finite-size effects       
peculiar to the $|Q|=1$ charge sector \cite{pm,overi,mnpNP}.                
For the generation of configurations, we use the Wilson action.

\section{Improved cooling and restricted improved cooling}

In \cite{mnp2,mnpNP} we introduced ``improved
cooling" (IC), based on a lattice action with improved scale invariance,
as a well defined method to systematically
eliminate UV noise and 
dislocations while preserving the topological charge
associated with instantons above a certain size threshold
$\rho_0 \sim 2.3a$.  
Here we shall recall some results of IC
that are relevant for our discussion.

IC uses an action whose 
instanton solutions are scale-invariant for sizes 
beyond the {\it dislocation threshold} $\rho_0$
of the order of the lattice spacing $a$. 
From a tree-level analysis \cite{overi} one can determine a five-loop combination
\bea
S_{m,n} &=& \ {1 \over {m^2 n^2}} \sum_{x,\mu,\nu} {\rm Re\ Tr}
\left( 1 - \ \ \ \ \twooneplaq  \hspace{1.5cm}\right) \nonumber\\
S&=&   \sum_{i=1}^5 c_i \ S_{m_i,n_i}.
\label{e.act1}
\eea
$(m_i,n_i) = (1,1),(2,2),(1,2),(1,3),(3,3)$ for $i=1,\ldots , 5$
 and:
\begin{eqnarray}
c_1 &=&(19- 55\  c_5)/9,\ \  c_2 =  (1- 64\  c_5)/9, \nonumber\\
c_3 &=&(-64+ 640\  c_5)/45,\ \  c_4 = 1/5 - 2\  c_5,
\label{e.act2}
\end{eqnarray}
which has no tree-level ${\cal O}(a^2)$ and ${\cal O}(a^4)$ corrections  
at any $c_5$. We have chosen $c_5=1/20$, which appears to minimize the
${\cal O}(a^6)$ corrections for instanton configurations. 

The cooling  
algorithm is derived from the equations of motion
\be
\label{eq}
U_\mu(x) W_\mu(x)^\dagger - W_\mu(x) U_\mu(x)^\dagger = 0 \; ,
\ee
where $W$ is the sum of staples connected to the link $U_\mu(x)$ in Eq.~(\ref{e.act1}).
For SU(2), for instance, 
cooling amounts to the substitution (for clarity we drop the  
indices when possible):
\be
U \rightarrow U' = V = W/||W|| \ , \ ||W||^2=\frac{1}{2} \Tr(W W^\dagger)\;.
\label{e.sta}
\ee
Above the {\it dislocation threshold}, $\rho_0 \simeq 2.3 a$, 
instantons are completely stable to
any degree of cooling (at least if they are not affected by finite-size
effects; see \cite{mnpNP} for a discussion of this point).
The corresponding
{\it improved charge density} using the same combination of loops
leads to an integer charge already after a few cooling
sweeps.

We now supplement this cooling procedure by imposing the  constraint
that only those links be updated, which violate the equation of motion by
more than some
$\tilde{\delta}$: 
\be 
U \rightarrow U' = V \qquad {\rm if} \qquad  
\tilde{\Delta}_\mu^2 (x)=  \frac{1}{a^6} {\rm Re} \Tr (W W^\dagger - 
(U W^\dagger)^2)\geq\tilde{\delta}^2,
\label{cond0}
\ee
where $\tilde{\Delta}_\mu (x)$ is the dimensionful 
square of the lattice equations of motion.
In the continuum limit (see \cite{map})
\be
\label{e.contl}
\tilde{\Delta}_\mu^2 (x) \rightarrow -2 \ 
\Tr((D_\nu F_{\nu \mu}(x))^2)\;;
\ee
hence we are locally requiring that the continuum equations of 
motion be satisfied to a degree 
specified by $\tilde{\delta}$.
This corresponds to a constrained minimization problem.
For $\tilde{\delta}\!=\!0$ 
cooling proceeds unrestricted, i.e. all links are
updated at every iteration, whereas for large values of $\tilde{\delta}$ no 
changes are made at all and the configuration remains uncooled.
Variation of $\tilde{\delta}$ then admits a smooth interpolation between these
extreme cases.

In particular, for small $\tilde{\delta}\!>\!0$ the 
configuration is changed only locally, where
it is far from a classical solution, but freezes where it 
has approached a 
solution to a certain approximation specified by $\tilde{\delta}$
through Eq.~(\ref{cond0}). 
Since the equations of motion
contain gradients of the fields, $\tilde{\delta}$ controls the energy 
of the fluctuations around classical solutions 
and acts as a filter for short wavelengths.
Hence cooling ends in a
configuration slightly above the minimal action, the additional 
``action"
being distributed smoothly over the lattice. We call this constrained 
variant of IC ``restricted improved cooling" (RIC).

 In this paper we have used a slightly different version of 
(\ref{cond0}), specific for SU(2): we replace (\ref{cond0})
by the condition:
\be \label{cond}
U \rightarrow U'  = V \qquad {\rm if} \qquad
\Delta_\mu(x)^2= \frac{1}{a^6} {\rm Tr}(1-U V^{\dagger})\geq\delta^2 \;,
\ee
with $V$ as in (\ref{e.sta}). In the continuum
limit $\delta \propto \tilde{\delta}$ \cite{map}.

\begin{table}[ht]
\begin{center}
\label{sat}
\begin{tabular}{|c|c|c|c|}
\hline
Lattice&{$a^3 \delta$}& $\delta$
  & Saturation \\
& & (fm$^{-3})$& at sweep No. \\ \hline
\hline
$a\!=\!0.12$ fm&0.08 & 46.30&5  \\ \hline
"&0.04  &23.15&  7 \\ \hline
"&0.02  &11.57& 11 \\ \hline
"&0.0085 &4.92& 22  \\ \hline
"&0.005  &2.89& 40  \\ \hline
\hline
$a\!=\!0.06$ fm&$0.04/8$&23.15  & 20  \\ \hline
"&$0.02/8$ &11.57 & 35  \\ \hline
"&$0.085/8$&4.92  & $\approx  85$  \\ \hline
"&$0.005/8$&2.89 & $\approx $ 180\\ \hline
\end{tabular}
\caption[]{{\it The number of cooling sweeps after which saturation is reached
for RIC.}}
\end{center}
\end{table}

\begin{figure}[tbp]
\vspace*{-1.0cm}
\centerline{\hspace{-2mm}
\epsfxsize=7.0cm\epsfbox{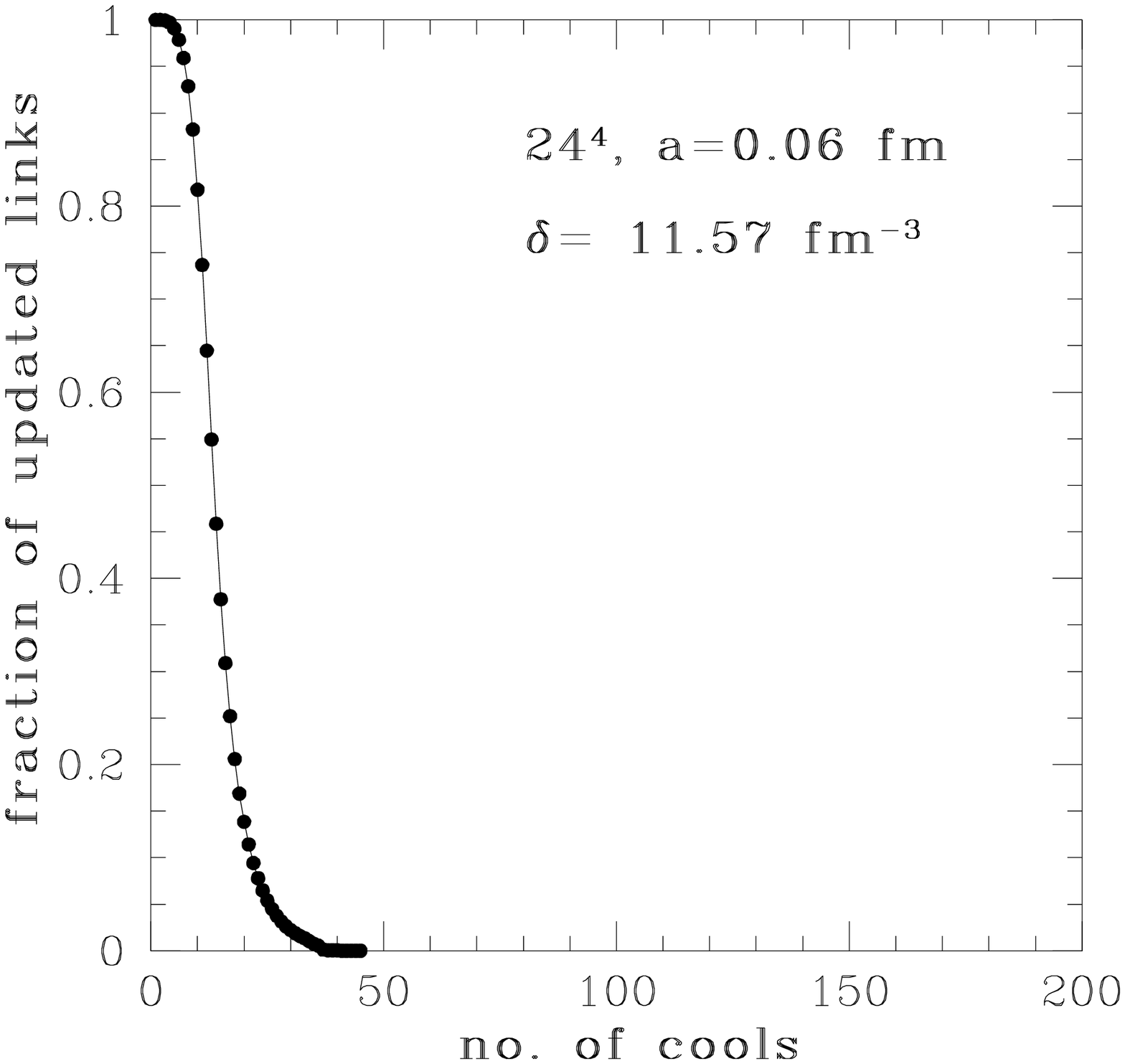}
\hspace{-.2cm}
\epsfxsize=7.0cm\epsfbox{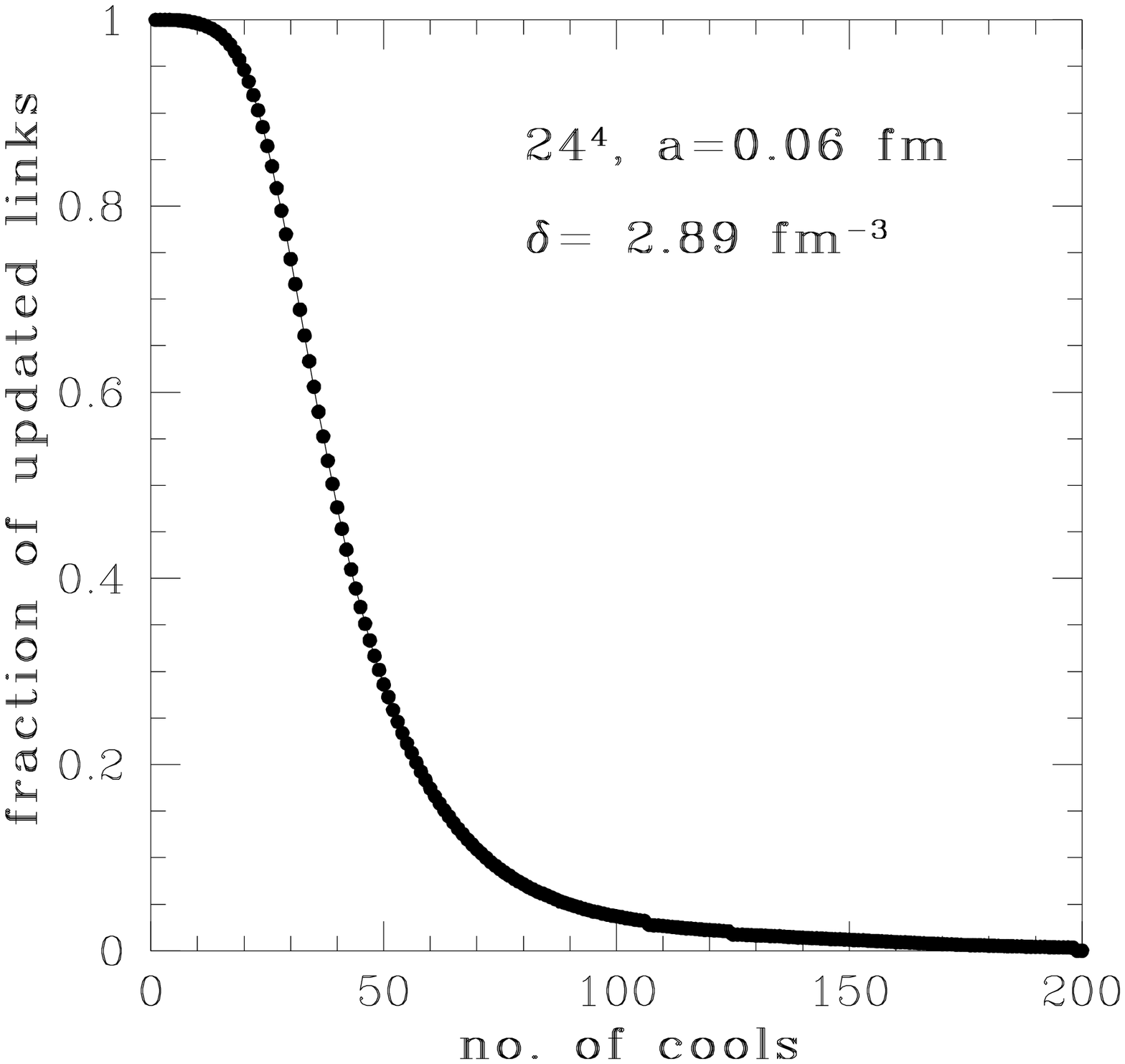}}
\caption[]{\label{f.sat}
{\it Fraction $\nu(n)$ of links updated per RIC iteration
as a function of $n$ for the $a=0.06$ fm
lattice.
}}
\end{figure}

Since RIC does not update links that are sufficiently close to a solution,
it changes fewer links after
every iteration until no further changes occur. 
On the MC configurations 
we define saturation to be reached
when less than $1\%$ of the links are updated in a sweep and
end the cooling there\footnote{We put this limit for practical reasons.
We  checked that it has little influence on the results: for very rough
configurations (large  $\delta$) some narrow instantons may still shrink under
cooling beyond this threshold; the effect is, however, at most
$1$--$2\%$ and disappears for smaller $\delta$.}.
The number of cooling sweeps required to reach
saturation is determined by $\delta$. Some example values,  for
our SU(2) lattices, are given in
Table \ref{sat}. 
In Fig. \ref{f.sat} 
the fraction $\nu(n)$ of links updated per cooling iteration is shown 
for the $a=0.06$ fm lattice
at two different values of $\delta$. 
The figure 
elucidates the difference between IC stopped at a certain number of sweeps 
and RIC. For the former, of course, 
$\nu$ is a step function whereas it is smeared out for RIC.

\section{Setting a scale for cooling}
\begin{figure}[tbp]

\vspace*{-2.2cm}

\centerline{\hspace{-.8mm}
\epsfxsize=8.5cm\epsfbox{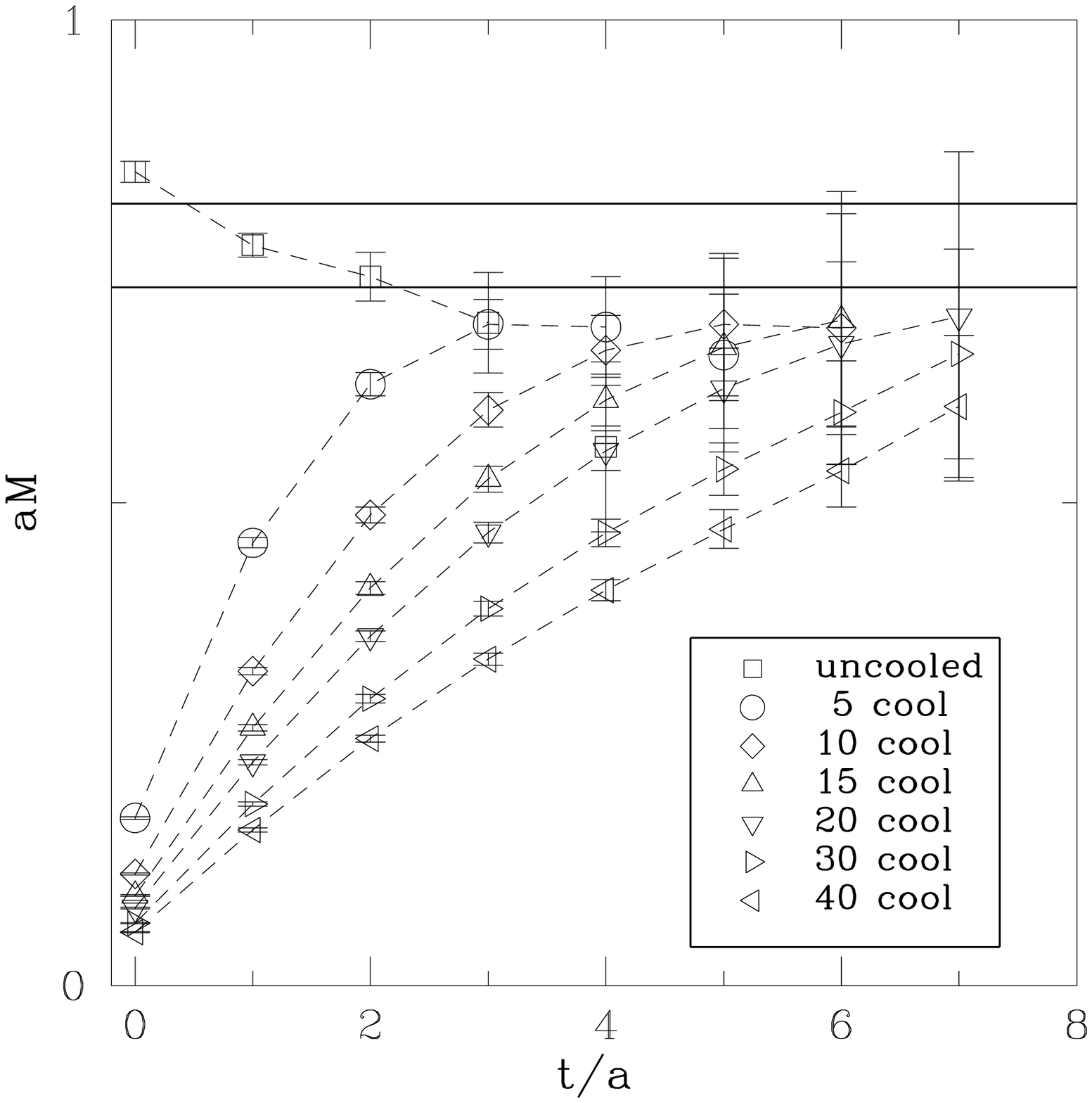}
\hspace{-.6cm}
\epsfxsize=8.5cm\epsfbox{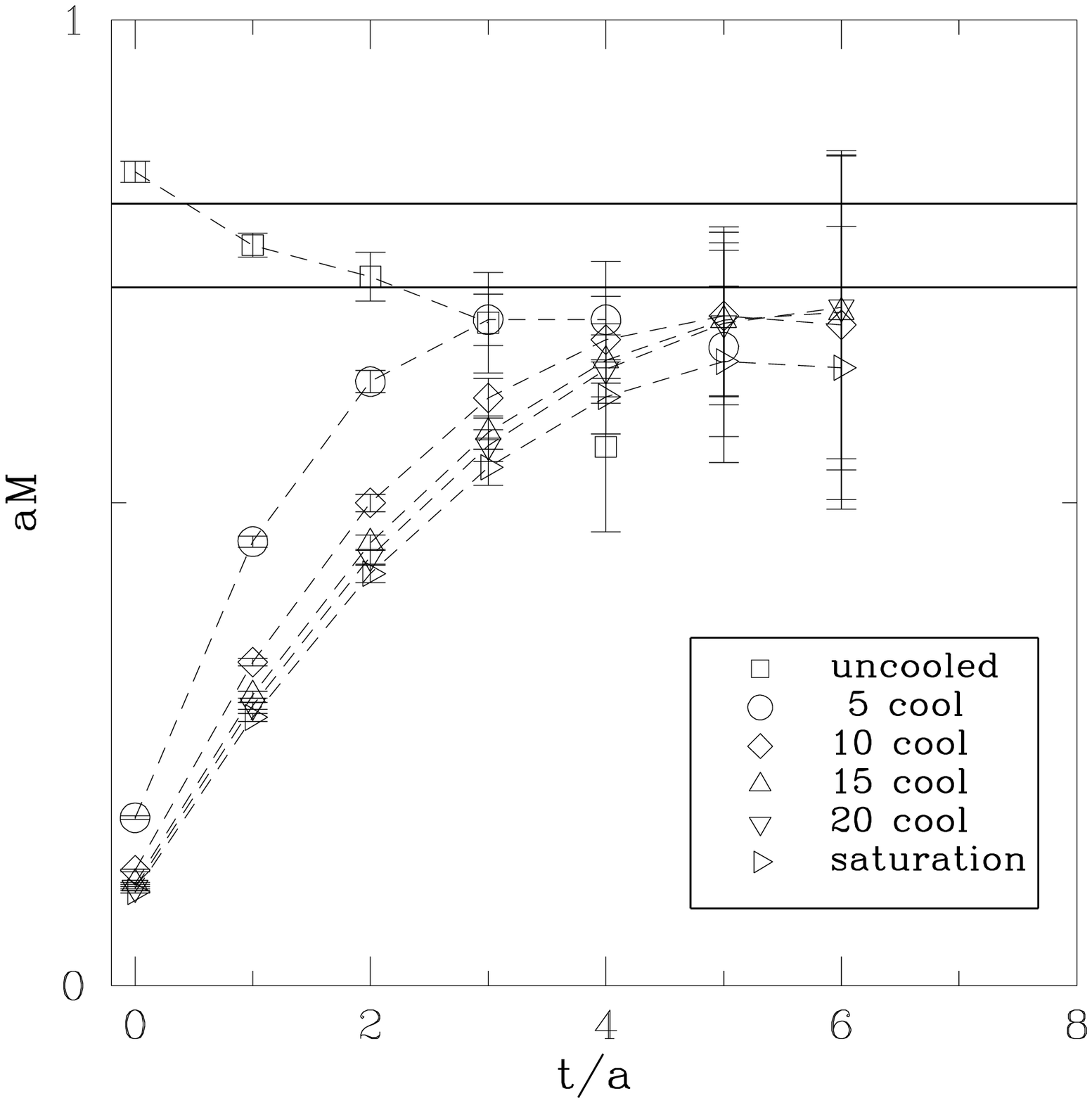}}


\caption[]{\label{cool}
{\it Behaviour of effective masses of the Polyakov loop under cooling.
     Left: IC.
     Right: RIC with $\delta=2.89$ fm$^{-3}$. The dashed lines are to guide 
the eye.
The horizontal line gives  a string tension value (within the corresponding
error band) as determined from \cite{mite}. 
}}
\end{figure}
\begin{figure}[tb]
\vspace*{-2.2cm}

\centerline{\hspace{-.8mm}
\epsfxsize=8.5cm\epsfbox{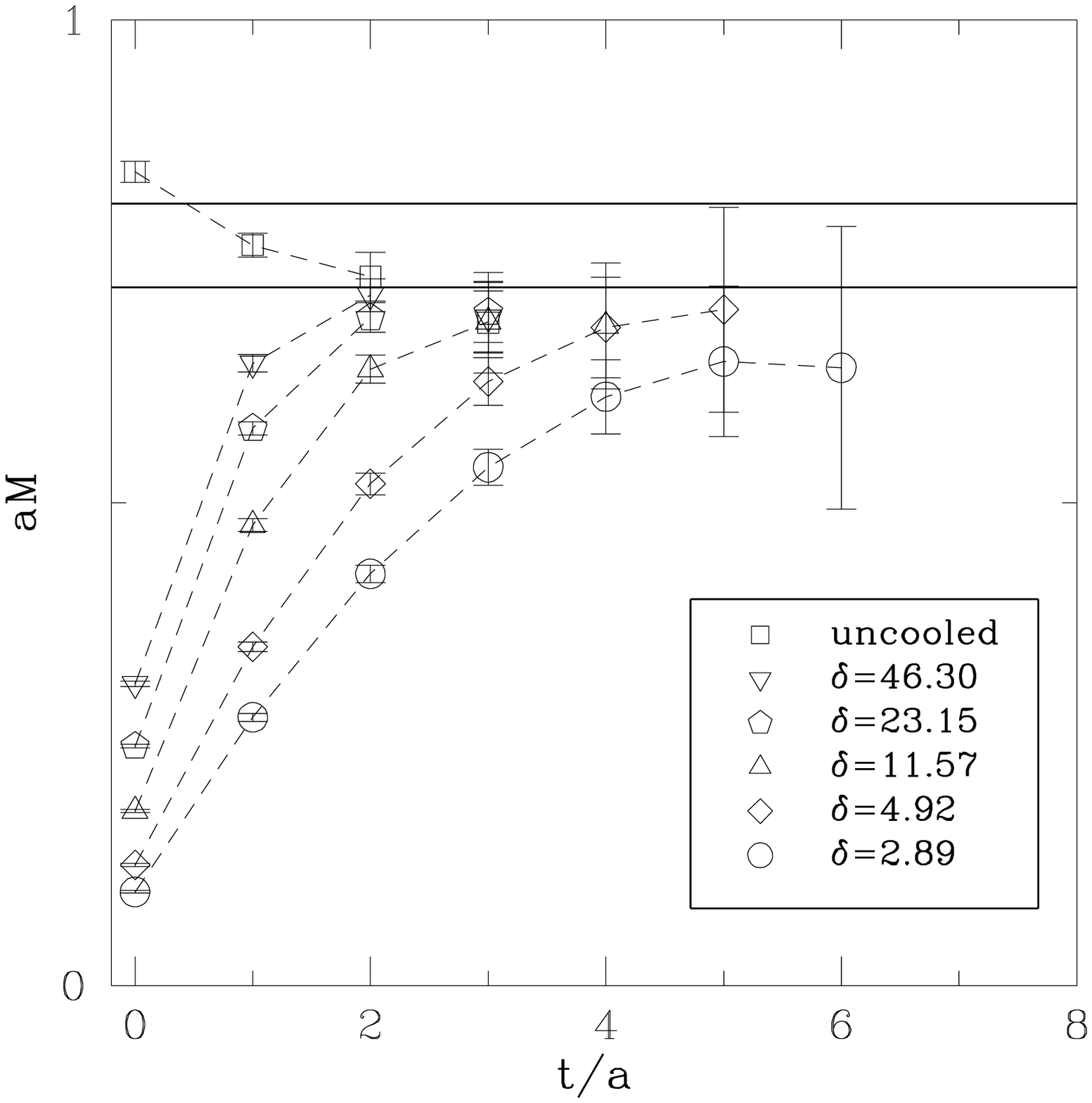}
\hspace{-.6cm}
\epsfxsize=8.5cm\epsfbox{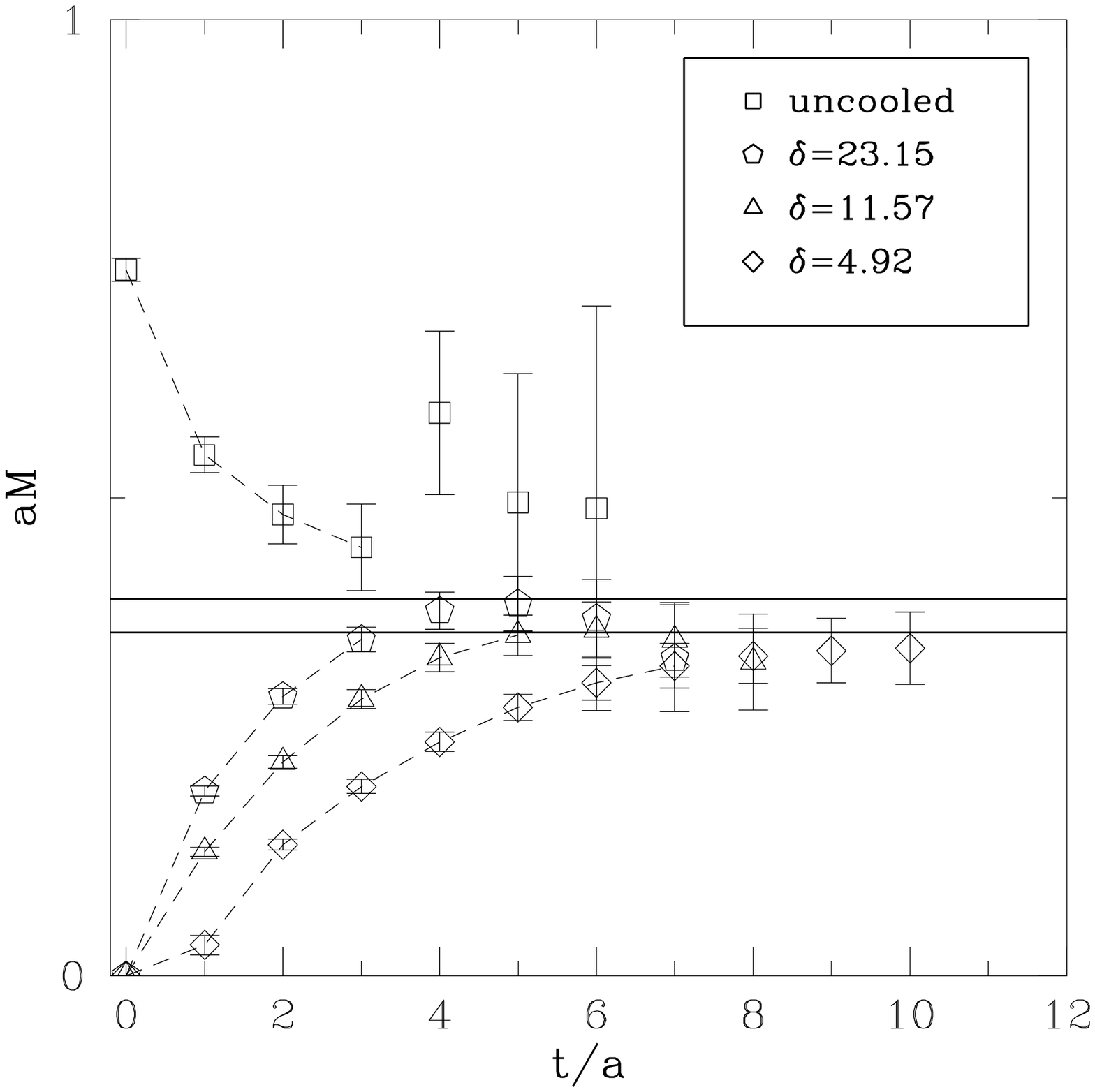}}


\caption[]{\label{delta}
{\it The cooling radius of RIC as a function of
$\delta$.  Left: $12^3\times 36$, $\beta=2.4$,\\ $a=0.12$ fm.
Right: $24^4$, $\beta=2.6$, $a=0.06$ fm.
}}
\end{figure}

In order to establish a physical scale up to which cooling is effective,
we study the behaviour of the string tension calculated on the cooled 
configurations. Such analyses have been carried out for other
variants of cooling in \cite{gon95,dig89}, indicating that the string tension 
is lost over increasingly large distances while cooling proceeds.
In RIC, however, the amount of cooling is related locally to
the equations of motion, which have a continuum limit.
Hence, we expect to be able to find a scaling criterion for tuning the 
cooling parameter $\delta$ 
and thus define it directly in terms of a physical scale.

A standard way to calculate the string tension is to extract it
from the exponential decay of the correlation function of spatial
Polyakov loops $P^{(L)}$ of length $L$ according to \cite{for85}
\be
\label{string}
  \sum_{\vec{x}}\,\left\langle P^{(L)}(x)P^{(L)\dag}(0)
                  \right\rangle \simeq e^{-a M_P^{(L)} t},\quad
   aM_P(L)=a^2\sigma_L L,\quad a^2\sigma_\infty = a^2\sigma_L+\frac{\pi}{3L^2}\;.
\ee
In order to see when the asymptotic value $aM_P(L)$ is reached, one
may look for plateaux in the effective mass on a given time slice,
\be
aM(t)=-\frac{1}{L} \ln \frac{P^{(L)}(t+1)}{P^{(L)}(t)} \;.
\ee
For uncooled configurations, the signal is too noisy to extract the
exponential fall-off. This problem is remedied by employing 
fuzzing techniques \cite{alb} to produce smeared operators with
better overlaps onto the ground state. For the cooled configurations,
this is usually not necessary as each cooling update includes the equivalent
of one fuzzing step.

In Fig.~\ref{cool} we show, for the $a=0.12$ fm
lattice in Table \ref{t.sim},
the effective mass of the Polyakov loop 
for various numbers of cooling sweeps using IC or RIC 
($\delta=2.89$ fm$^{-3}$). 
Since in this exploratory study we have rather poor statistics, we do not
try to obtain a precise value for the string tension but are only interested
in comparing the plateaux obtained from the 
uncooled and cooled correlation functions. 
For illustration we indicate in the figure 
the bandwidth of the value obtained from other data sets for the 
string tension \cite{mite}.
The previously reported effect of a diminished
string tension through cooling 
\cite{gon95,dig89} is clearly visible, and
in both cases there is a ``cooling radius'' at which the effective masses
merge with their counterparts from uncooled configurations, 
and beyond which the string tension
is still unaffected.
However, in the case of IC (as well as other cooling variants), 
this cooling radius moves to larger distances
with the square root of the number of iterations \cite{pisa1}.
As a consequence, 
the physical properties
of the system are changed up to distances that are so large  that
beyond them any measurement of physical observables
with reasonable statistical errors is practically impossible.
In contrast, RIC saturates in the sense that no more changes take place 
above a certain number of iterations. 
The physical properties of the system beyond the cooling 
radius reached at saturation appear to remain intact.

Rather than by the number of iterations,
the cooling radius for RIC
is determined by the choice of the parameter $\delta$, as displayed 
in Fig.~\ref{delta}. A larger choice for $\delta$ results in
an  ``earlier'' saturation, in the sense that links are
left untouched when they represent a cruder approximation
to the equations of motion, cf. Eq.~(\ref{eq}). As elucidated in the figure,
this results in a smaller cooling 
radius.
 The clear approach to saturation  
offers the possibility to fix the desired cooling radius {\it before}
a simulation and eliminates the need to monitor the cooling process,
which so far had to be stopped ``by hand'', based on a 
subjective judgement on the trade off
between smoothness of configurations and conservation of physical properties. 

Furthermore, since the parameter $\delta$ is expected to have
a continuum limit, it should be possible
to establish a systematic mapping between the values of $\delta$ for a 
given lattice
spacing and a physical length scale, up to which the smoothing
process is active. This would permit a clean separation of physical scales into
an ultraviolet sector, where smoothing is active, and an infrared sector,
in which all physical properties are preserved.
The naive expectation is that, close to the continuum limit, 
the dimensionful quantity $\delta$ should scale, cf. eq.~(\ref{e.contl}).
To check this behaviour we have repeated this study on the $a=0.06$ fm lattice
in Table \ref{t.sim},  which 
halves the lattice spacing of the previous one while
preserving the same 3-dimensional volume. The results 
for the effective mass on this lattice are presented in Fig.~\ref{delta} (right).
Despite the fact that the statistics is not very large, there is
clear confirmation of this expectation. 
We extract the cooling radius $r(\delta)$ from Fig.~\ref{delta}
as the value of $t$ at which the effective masses merge with the uncooled
results and the string tension is recovered. 
A comparison of the data at fixed $\delta$ 
for the two  values of $a$ shows (at the level of precision of these data) 
good scaling properties   
and therefore allows us to estimate the physical cooling radius to be

\be
r(\delta) \approx 0.8 \ \delta^{-1/3} \; .
\label{e.cscale}
\ee

Since the string tension gives the fundamental physical scale of the theory,
we {\it conjecture} that {\it the cooling radius defined in this way is valid
for all physical observables}, in the sense that 
all physical 
properties defined on scales larger than $r(\delta)$ can be measured
on configurations RI-cooled with $\delta$. In the following we shall apply
this conjecture to the observation of the topological structure.

\section{Instantons and RIC}

In order to provide a reliable extraction of the topological information from the
MC configurations, there are two properties a smoothing algorithm should have: 

(1) It should allow a clean separation between dislocations (small instantons at
the scale of the cut off) and physical instantons.  
The former should be eliminated under smoothing, the latter should remain 
unaltered.

(2) Once we have fixed a physical distance below which smoothing is active,
it should guarantee that the structure above it, among others instanton--anti-instanton
pairs, remains unchanged.

Improved cooling was designed to achieve (1) and we shall prove below that RIC 
preserves this property. With respect to the string tension,
RIC also satisfies requirement (2). As we shall see this is also the case 
for I-A pairs.

We present below the study of a set of prepared small instantons and I-A pairs
that give an indication on the performance of RIC with respect 
to these two points. 
The reader interested only in the Monte Carlo results may skip this section and
go directly to section 5. 

\subsection{Dislocations and RIC}

 The improved action in Eqs. (\ref{e.act1}), (\ref{e.act2}) was designed
to minimize violations
of scale invariance: it is almost independent of the instanton size $\rho$
when the latter is above the threshold $\rho_0\approx 2.3a$.
Instantons with $\rho > \rho_0$ are practically left unchanged
under cooling, irrespective of whether we use IC or RIC. 

\begin{figure}[tbp]

\vspace*{-2.0cm}

\centerline{\hspace{-2.5mm}
\epsfxsize=8.2cm\epsfbox{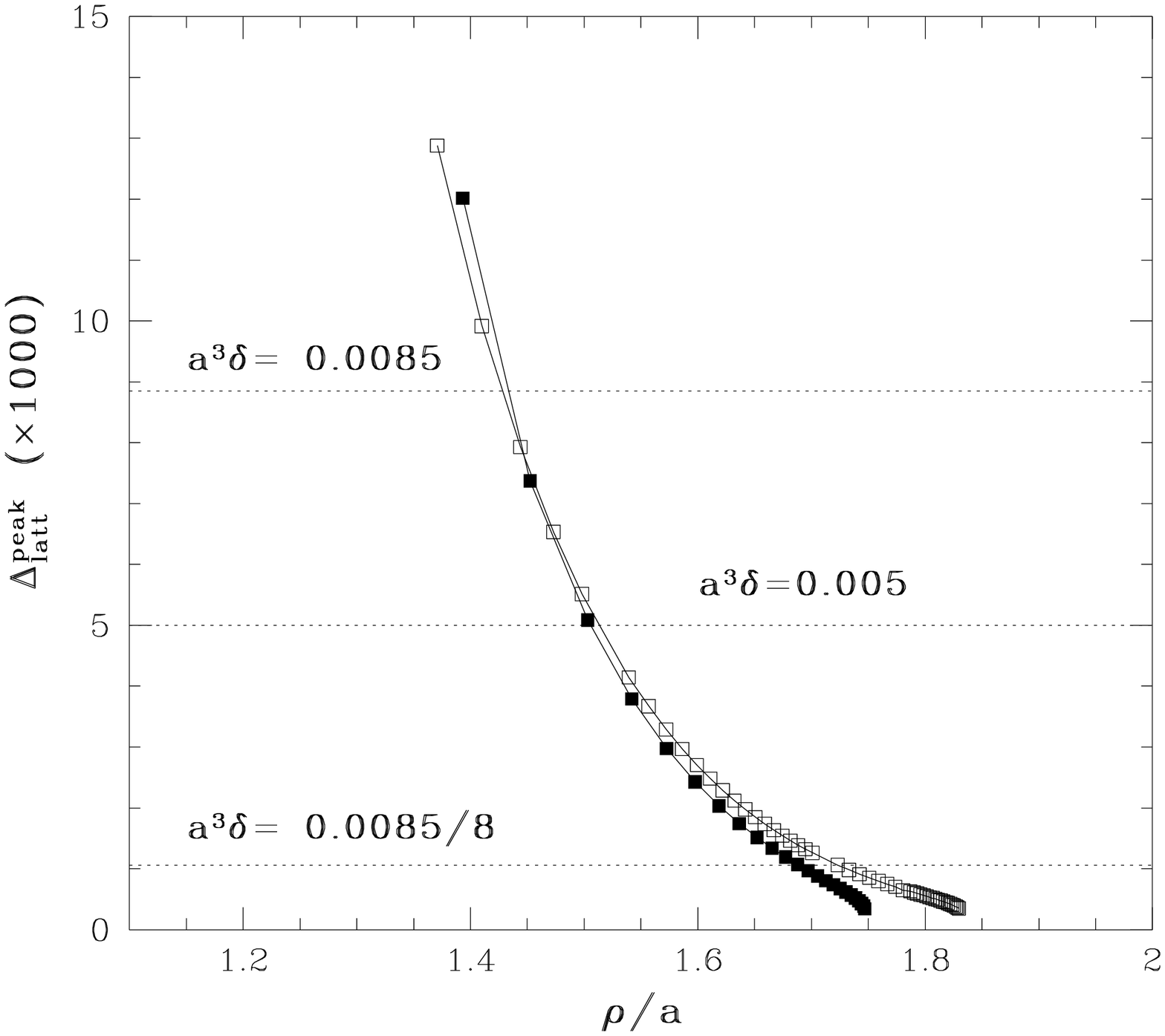}
\hspace{-.2cm}
\epsfxsize=8.2cm\epsfbox{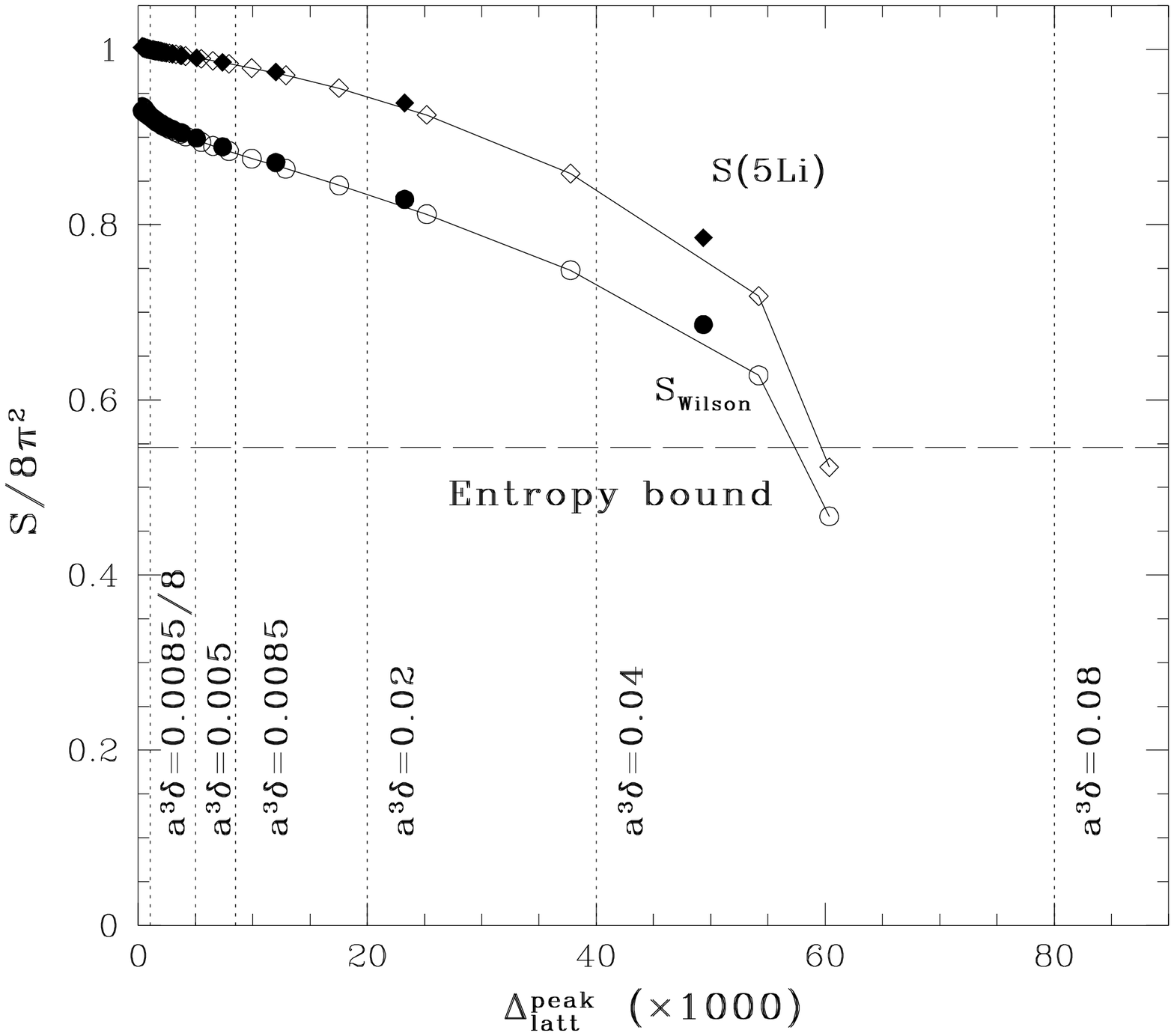}}
\caption[]{\label{f.actr}
{\it  Left:  Dimensionless deviation from a solution of the equations of motion:
$\Delta^{\rm peak}_{\rm latt}$ -- see Eq. (\ref{eq.small}) --
as a function of the instanton size in lattice units.
Right: Action of small instantons as a function of $\Delta^{\rm peak}_{\rm latt}$.
The horizontal dashed line indicates the entropy bound:
instantons with $S_{\rm Wilson} $ below this line should be considered
as dislocations \cite{pt}.
On the left (right) figure the horizontal (vertical) dotted lines show
 the values of $a^3 \delta$ used in our RIC simulations.
Open (filled) symbols correspond to the $a=0.12$ $(0.06)$ fm lattice.
 }}
 \end{figure}

The situation is, however, quite different
for small instantons with $\rho<\rho_0$. Under IC these
instantons shrink and disappear through the lattice ``holes". Such small
instantons are expected to deviate considerably from solutions to the equations
of motion. To check this conjecture we have evaluated
this deviation for a set of instantons with size below $\rho_0$.
We present on Fig. \ref{f.actr} (left)  the dependence on
$\rho/a$ of the lattice dimensionless quantity:
\be
\Delta^{\rm peak}_{\rm lat} =  \frac{a^3}{4} \sum_\mu \Delta_\mu(x^0),
\label{eq.small}
\ee
with $\Delta_\mu(x)$ from Eq. (\ref{cond}), evaluated on top of
the instanton. As expected, for $\rho < \rho_0$ the deviation from a solution
increases
in a rather steep way with decreasing instanton size.
RIC limits $\Delta_\mu(x)\ \le \delta,\ \  \forall x$;
in the figure we indicate by horizontal dotted lines the values of
$a^3\delta$ used in our RIC simulations. Instantons with 
$\Delta^{\rm peak}_{\rm lat}$ above each line will be destroyed under
RIC with the corresponding $\delta$. 
We may as well say that instantons with size below $\rho_0^{RIC}(a^3\delta)$
will shrink and disappear under RIC with $\delta$, where
$\rho_0^{RIC}(a^3\delta)$ is given by the intersections of the curves
on Fig. \ref{f.actr} (left) with the horizontal lines.

We present on the figure data for both $a=0.12$ and 0.06 fm.
It is important to note that we are plotting the deviation from a solution
evaluated in lattice (not physical) units.
The rather weak dependence on $a$ for $\rho/a <2$ indicates that for
instantons close to the lattice spacing the scaling law given
by Eq. (\ref{e.contl}) does not hold, and the deviation from a solution 
is fixed solely in terms of the size in lattice units, i.e. $\rho/a$.

To summarize, setting $\delta$ allows to fix a threshold
$\rho_0^{RIC}(a^3\delta)$ on the size of the instantons that will be preserved under RIC.
$\delta\!=\!0$ corresponds to the IC threshold
$\rho_0\equiv\rho_0^{RIC}(\delta\!=\!0)\approx 2.3a$.
For $\delta \ne 0$, the dislocation threshold (in lattice units) is smaller
for larger lattice spacing. 

As pointed out before, the existence of this small instanton threshold 
is important to avoid the presence of dislocations after cooling. 
A semi-classical argument due to Pugh and Teper \cite{pt} indicates
that lattice instantons with action below the so-called
entropy bound, $S_{\rm lattice}(\rho) \le 48\pi^2/11$, dominate the path integral
in the continuum limit and give rise to the lattice
artefacts known as dislocations; for details of the argument, see \cite{pt}.
Since we have generated the MC configurations with the Wilson action ($S_{\rm Wi
lson}$)
this is the relevant quantity to characterize small instantons as dislocations.
 In  Fig. \ref{f.actr} (right) we show  the improved action as well as $S_{\rm Wilson}$
 for the instantons in Fig. \ref{f.actr} (left).
According to the above, instantons with $S_{\rm Wilson}$ below the entropy bound
(indicated on the figure by the dashed horizontal line)
should be considered as dislocations. As can be seen from the figure
the deviation from a solution of the equations of motion for
such configurations is quite large. It is then obvious that they can be eliminated
by putting a threshold on the maximally allowed deviation and this
is precisely what RIC does.
We indicate in the figure by dotted vertical lines the values
of $a^3\delta$ used in our RIC simulations. 
It seems that topological lattice artefacts are under control with RIC as long 
as we keep $a^3\delta < 0.04$.

\subsection{I-A pairs and RIC.}

 One of the motivations for RIC was to introduce a physical scale
above which the physical information concerning I-A pairs remained unaltered. 
Both on the lattice and in the continuum, 
these configurations are not solutions of 
the classical 
equations of motion, unless the instanton and anti-instanton are infinitely 
separated from each other.
Since ordinary cooling is an unconstrained local minimization of the action,
when cooled with it the I and A annihilate and decay into the
trivial vacuum. The information concerning I-A pairs extracted from ordinary
cooling is hence distorted in an uncontrolled way. 
RIC cooling, on the contrary, will leave unchanged those pairs for which
the deviation from a solution remains everywhere smaller than $\delta$.
The expectation is that this provides a means 
to fix a physical scale above which the information concerning pairs
is unchanged. 

\begin{figure}[tbp]

\vspace*{-2.0cm}

\centerline{\hspace{-2mm}
\epsfxsize=8.2cm\epsfbox{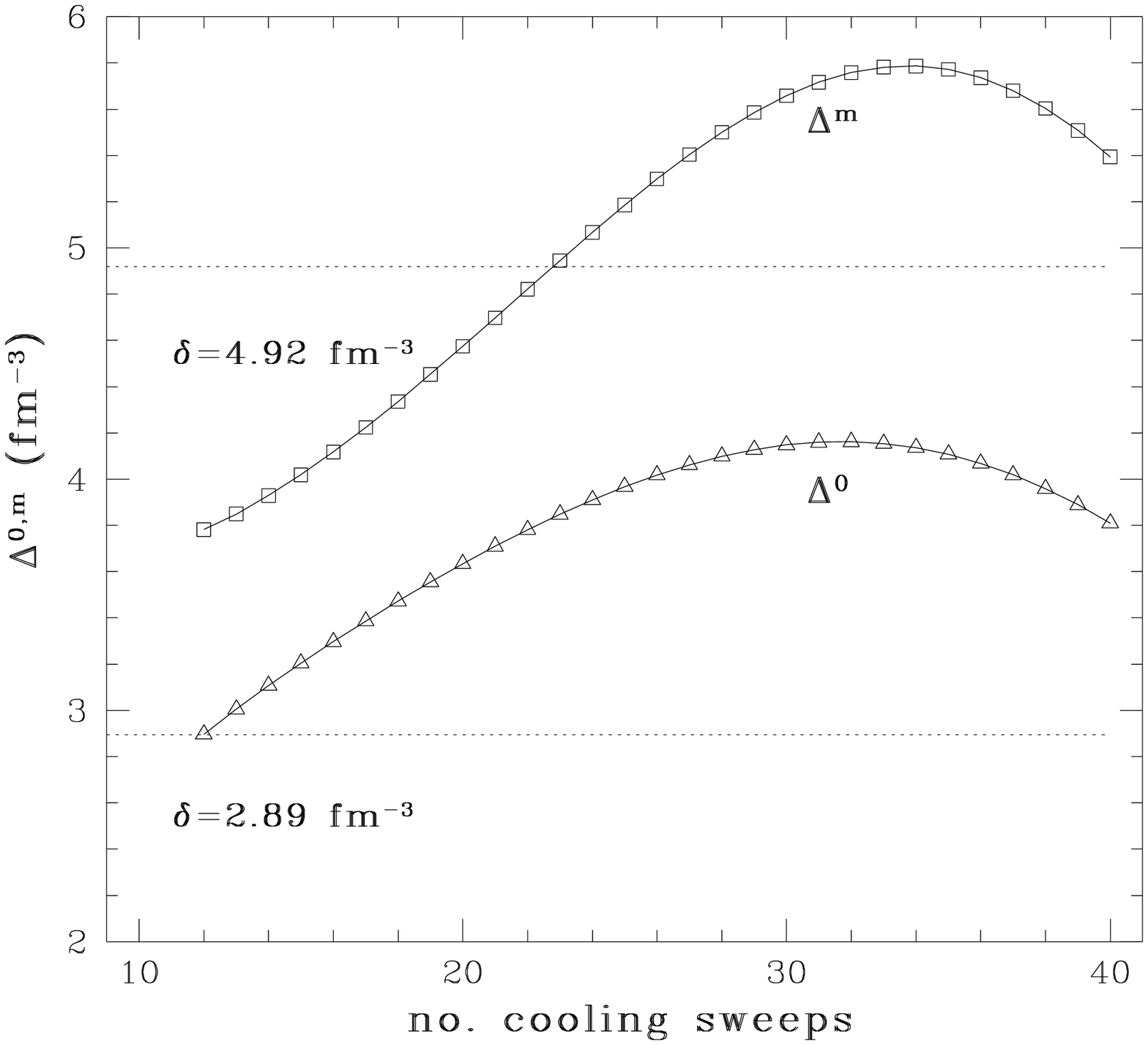}
\hspace{-.2cm}
\epsfxsize=8.2cm\epsfbox{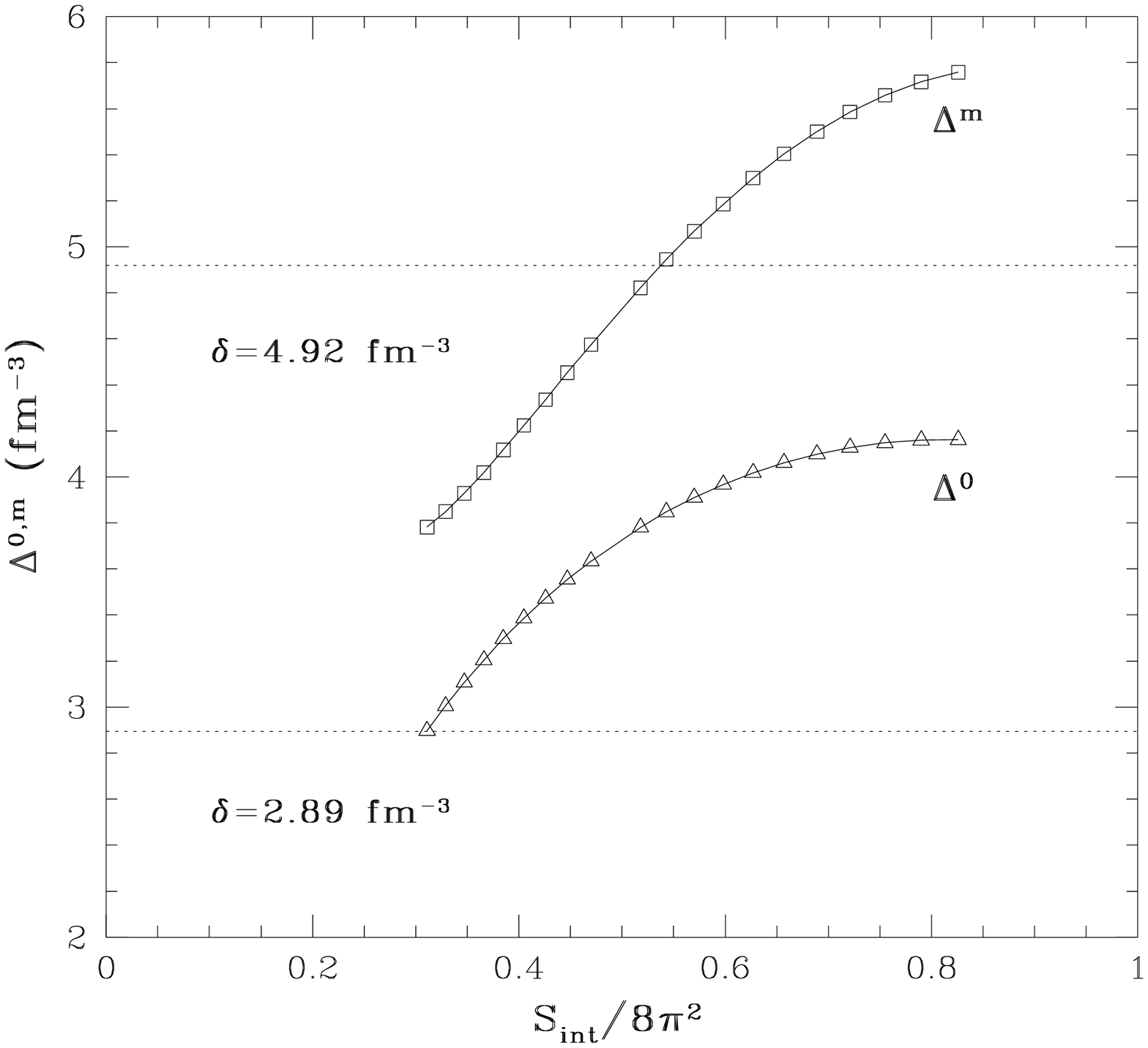}}
\caption[]{\label{f.ia} {\it
Left:
Evolution of the deviation from a solution of
the equations of motion $\Delta^{0,m}$ -- 
see Eq. (\ref{e.disia}) -- under IC
as a function of the number of cooling sweeps.
Right:
$\Delta^{0,m}$ vs $S_{\rm int}^{IA}$ -- see Eq. (\ref{e.sint}).
The results are obtained on the $a=0.12$ fm lattice.
The horizontal dotted lines show the values
of $\delta$ used in our simulations.
Pairs for which $\Delta_\mu(x) $
is below the line for all $x$
are preserved under RIC with $\delta$.
}}
\end{figure}

To test this hypothesis we have analysed a prepared set of I-A pairs
(similar studies for other variants of cooling have been performed in \cite{alvaro}).
 As in \cite{alvaro} we create a pair by putting together a 
$Q=1$ instanton and, at a distance $d_{IA}$ along the time direction,
the anti-instanton obtained from it by time reversal.
This configuration is I-cooled and its evolution is studied under cooling.
We have monitored in particular the change of the 
deviation from a 
solution as evaluated at the instanton centre ($x^0$) and at the 
mid-point between I and A ($x^m$):
\be
\Delta^{0,m}= \frac{1}{3}\sum_{i=1}^3 \Delta_i(x), \  {\rm with} \   x=x^0, x^m
\;,
\label{e.disia}
\ee
where $\Delta_i(x)$ is given by Eq. (\ref{cond}) (the deviation measured on links along the separation axis is much smaller,
therefore we did not include it in the estimate of $\Delta^{0,m}$).
The results are presented in Fig. \ref{f.ia} (left). 
This figure should be interpreted as providing a series of I-A
configurations labelled by the number of IC sweeps.
As expected, with proceeding IC $\Delta^{0,m}$
increases (the configuration gets distorted away from a solution)
until it decays into a trivial one; after that we can no longer recognize 
I and A as peaks in the energy density, the configuration  approaches the trivial
vacuum and $\Delta^{0,m}$ decreases again.
This picture changes under RIC. We have plotted in the figure
horizontal dotted lines corresponding to the 
values of $\delta$ used in our simulations. 
Whenever a configuration satisfies  $\Delta_\mu(x) \le \delta$ everywhere,
in particular at $x^0$ and $x^m$, it will be left untouched by 
RIC, i.e. the pair will be preserved and will not annihilate under 
RIC with $\delta$. Pairs with $\Delta^{0,m}$ above the line will, however, 
be destroyed.

 Our objective is now to fix which pairs are preserved, using a 
physical length scale, since this may allow us to relate their 
effects to the physical quantities characterized by that scale. 
As already mentioned, since I-A pairs are 
continuously connected with the trivial vacuum, there is an ambiguity 
(in the continuum) in distinguishing a pair
from a trivial fluctuation. The threshold separating the two is usually 
rather arbitrary and set ``by hand". One possibility is to do it
through the value of the interaction
between I and A,
\be
S_{\rm int}^{\rm IA}=16\pi^2-S^{IA}.
\label{e.sint}
\ee
It seems only justified to speak of an I-A pair instead of a fluctuation if 
$S_{\rm int}^{\rm IA}$
remains considerably smaller than the action of a ``non-interacting" pair: 
$16\pi^2$.
We can thus parametrize I-A pairs in terms of how small $S_{\rm int}^{\rm IA}$ is,
and use it as a threshold for which pairs will be considered
as physical.
We have monitored $S_{\rm int}^{IA}$ for the pairs in Fig. \ref{f.ia} (left) and
the results are presented in Fig. \ref{f.ia} (right). Whenever
we are able to still recognize the I and A there is a one to one
correspondence between the distance to a solution (notice
$\Delta^{0,m}$ is a dimensionful quantity) and $S_{\rm int}^{\rm IA}$.

If we parametrize I-A pairs through how small $S_{\rm int}^{\rm IA}$ is,
it is clear that setting $\delta$ is equivalent to putting a cut
on which pairs will be considered as physical and which will be thrown away
according to the value set for  the maximal $S_{\rm int}^{\rm IA}$. 
It is in this sense that RIC implements the valley method constraint, since it
amounts to a local constrained minimization of the action that prevents
the interaction to act along the valley between I and A,  
if such an interaction is smaller than the chosen $S_{\rm int}^{\rm IA}$.

\section{Monte Carlo results}
\subsection{Topological susceptibility from RIC}

\begin{table}[ht]
\begin{center}
\begin{tabular}{|c|c|c|c|c|c|c|}
\hline
Lattice&$a^3\delta$&$\delta$ fm$^{-3}$
& $^4\sqrt{\chi}$ MeV
& $\langle Q^2 \rangle$
& $\langle S \rangle/8\pi^2$
& $\langle |Q| \rangle$
\\
\hline
 $a\!=\!0.12$ fm& no cooling& &172(2)&7.4(3) &10634(1)&2.16(5) \\

\hline
"&0.08&46.30&190(3)&10.9(7) &443.9(2) &2.6(1)\\
\hline
"&0.04&23.15&194(4) &11.9(9)&210.0(2) &2.8(1)\\
\hline
"&0.02&11.57&199(3)&13.2(8) & 109.7(2)&2.93(9)\\
\hline
"&0.0085&4.92&198(2)&13.0(6) & 51.7(1)&2.91(6) \\
\hline
"&0.005&2.89&197(3)&12.6(8) &33.3(3) &2.85(8)\\
\hline
"&IC nc=20&&199(2)&13.1(6) & 25.2(2)&2.91(8)\\
\hline
"&IC nc=40&&197(2)&12.6(7) & 15.4(2)&2.86(8)\\
\hline
\hline
 $a\!=\!0.06$ fm&no cooling&&297(4)&22(1) & 50482(4)&3.7(1)\\
\hline
"&0.04/8&23.15&203(7)&4.7(6) &80.4(2)&1.7(1)\\
\hline
"&0.02/8&11.57&203(7)&4.8(5) &40.3(2)&1.7(1)\\
\hline
"&0.0085/8&4.92&210(7) &5.4(5) &18.8(2)&1.8(1)\\
\hline
"&IC nc=20&&205(7) &4.9(7)&30.8(2)& 1.7(1)\\
\hline
"&IC nc=50&&205(7) &5.0(7)&12.9(2)&1.7(1) \\
\hline
\end{tabular}
\caption[]{\label{t.all}
{\it Topological susceptibility $\chi$, average charge squared
$\langle Q^2 \rangle$, action $\langle S\rangle/8\pi^2$ and absolute 
value of the topological charge $\langle |Q| \rangle$, from
 IC and RIC.
}}
\end{center}
\end{table}

\begin{figure}[tbp]
\vspace{8.0cm}
\includegraphics{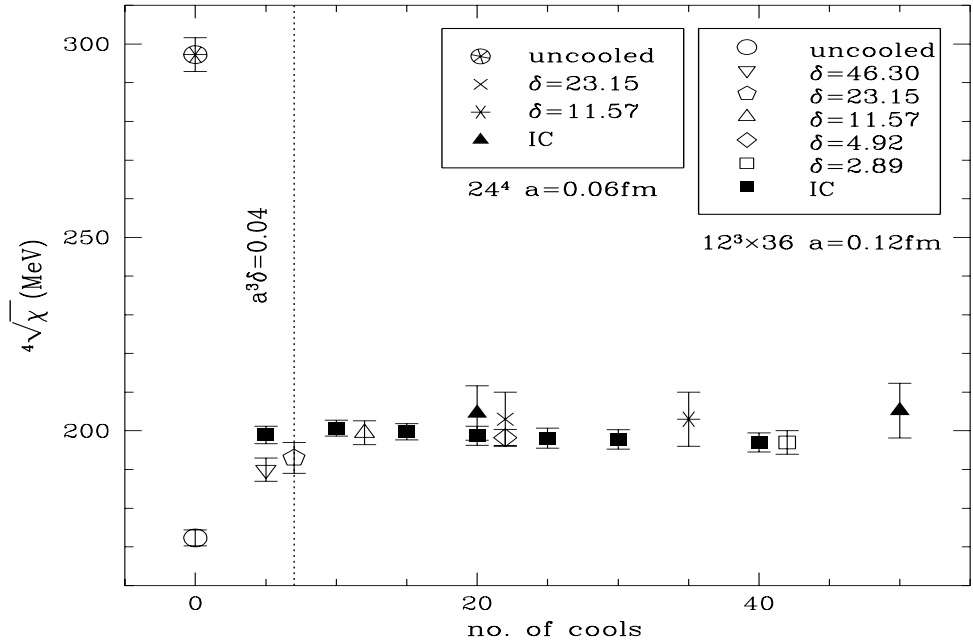}
\caption[]{\label{f.chi}
{\it Topological susceptibility with IC and RIC.
The results from RIC are plotted at the number of cooling sweeps
corresponding to saturation -- see Table \ref{sat}. 
}}
\end{figure}

The results for the topological susceptibility are presented in Table
\ref{t.all} and Fig. \ref{f.chi}.
 For comparison we also include in the
table and the figure the results for the topological susceptibility extracted
directly from the MC configurations (without smoothing). 
As expected, the raw Monte Carlo data
indicate strong deviations from scaling which, comparing to the IC
results, are particularly
severe for the finer lattice \footnote {This is to be expected for
MC configurations
generated with the Wilson action, since the simple semi-classical argument
quoted above
\cite{pt} indicates that the contribution of dislocations
to the Wilson action partition function diverges in the continuum limit.}.

 The expectation expressed in section 4.2 that changing $a^3\delta$ 
allows to change
the threshold on small instantons, and hence on dislocations, is corroborated
by the MC data. The result on the $a=0.12$ fm lattice at $a^3 \delta = 0.08$ 
is clearly affected by dislocations, as seen in the tendency to approach the
uncooled results. According to this observation and in view of the 
strong scaling violations of the uncooled data, it seems quite dangerous
to extract the value of the topological susceptibility by extrapolating to
zero smoothing. Generally, a minimal amount of smoothing may be indispensable 
to obtain a physically meaningful value for the
susceptibility; see also  \cite{erha}.

 By contrast, in the ``safe" region ($a^3\delta < 0.04$) we observe 
good agreement between the different results, including good scaling behaviour
when comparing the $a=0.12$ and 0.06 fm lattices. This is an indication
that most of the relevant physical instantons have been correctly taken into
account and have sizes above the small instanton thresholds 
$\rho^{RIC}_0(a^3 \delta < 0.04)$.

The results agree well with the Witten-Veneziano formula and with
results obtained previously by us using IC \cite{mnp2,mnpNP},
as well as by other groups using different methods  \cite{digiat,ts98,alvaro}.
Differences with the results in \cite{boulder} appear only for the 
largest $\beta$ 
value and amount to about 10$\%$ of the value of $\chi$.

\subsection{Instanton ensemble from RIC}

A first impression of the effect of RIC is already given by
the average value of the action presented in Table \ref{t.all}.
 Assuming that
after some smoothing the action is
mostly saturated by $|Q|=1$ instantons, the average action gives
an estimate of the total number of instantons.
It is clear that, under the above assumption,
large $\delta$ keeps more instantons, i.e.
less pairs are annihilated. Moreover the scaling behaviour of $\delta$
is confirmed within at most 10$\%$ difference by comparing the average value
of the action per unit volume between the $a=0.12$ fm lattice at
$\delta= 23.15$, $11.57$, $4.92$ fm$^{-3}$: $S/V=16.28(2)$, $8.78(2)$, $4.57(1)$
fm$^{-4}$;
and the corresponding values on the $a=0.06$ fm lattice:
$S/V= 18.49(4)$, $9.27(4)$, $4.32(4)$ fm$^{-4}$.

To associate the action with an instanton ensemble 
one has to extract in some way the
instanton content of the configurations. This is a difficult
task due to several reasons:

(i) as  pointed out before, there is an ambiguity in
distinguishing between an I-A pair and a trivial fluctuation;

(ii) this ambiguity is enhanced if the ensemble is dense, as seems to be the case,
since then the objects in the ensemble are strongly overlapping.

This last point sheds quite some doubts on the reliability and usefulness  
of the instanton parametrization, unless one can justify the introduction of 
some physical criterion to select the relevant pairs for the
process under consideration. 
 In particular, the results obtained 
for instance for the size distribution seem to be quite dependent 
on the particular method used to smooth and analyse the configurations
(see \cite{Neg}). 

Our particular way of extracting the instanton information
out of the configurations relies on two assumptions:
(1) instantons should appear as local self-dual peaks in the action and charge
density; (2) only pairs with $S_{\rm int}^{\rm IA}$ 
considerably smaller than $16 \pi^2$ 
should be considered as such.  Hence we approximate the action and charge
density by a superposition of self-dual non-interacting instantons and 
anti-instantons parametrized through the 1-instanton BPST ansatz
(further details on the procedure are provided in the Appendix). 
We can measure the departure of the real action and charge density from 
the above non-interacting ansatz through the quantities
\be
\varepsilon_q = \sqrt{
{{\int d^4x |q(x) - q_{fit}(x)|^2} \over {\int d^4x |q(x)|^2}}}
\label{e.dfit}
\ee
 and $\varepsilon_s$, defined as $\varepsilon_q$ with the replacement
$q\rightarrow s$.
In previous work we have adopted the prescription of performing a minimal
amount of cooling such that the deviations from the ansatz described above
were at most 0.3. Under this restriction, and using IC, we observed:

(i) for the minimal level of smoothing, a rather dense 
ensemble with a number of instantons by about a factor of 2 larger than the
dilute gas prediction $\langle N\rangle = \langle Q^2\rangle$, with $N$
the number of instantons
plus anti-instantons;

(ii) a size distribution peaked at about 0.4 fm and rather stable under further cooling
(a much smaller average size of
$\sim 0.2$ fm has been reported in \cite{boulder,boulder0}, extracted from 
smeared configurations and under extrapolation to no-smearing; see
5.2.2 for further discussion);

(iii) a rather homogeneous distribution of I and A, which seem to be randomly
distributed over the lattice.

For the largest values of $\delta$ in our RIC analysis
we will present results corresponding to values of $\varepsilon_{s,q}$ up to
0.6, where only a small part of the  action and charge density is 
correctly described by our instanton ansatz. These results should be 
taken with care, since they correspond to rather dense ensembles and,
as mentioned above, in such case there is a significant ambiguity in 
the extraction of the instanton content. The aim in the present work is
to study the results obtained by our instanton finder 
as a function of $\delta$ in RIC.  

\subsubsection{Results}

Now we turn to the analysis of the Monte Carlo configurations.
Results are presented in Table \ref{t.res}.

\begin{table}[ht]
\begin{center}
\begin{tabular}{|c|c|c|c|c|c|c|c|c|c|}
\hline
a &$\delta$ (fm$^{-3})$
& $r(\delta)$ 
& $\langle N/V \rangle$
& $\langle O_{II}\rangle$
& $\langle O_{IA}\rangle$
& $\langle d_{II}\rangle$
& $\langle d_{IA}\rangle$
& $\varepsilon_q$
 ($\varepsilon_s$)
& $\langle \rho\rangle$
\\
\hline
\hline
$0.12$&23.15  & 0.24(5) &14.86(7)&0.95&1.28&0.39&0.28
&0.75(0.55)&0.38\\ \hline
"           &11.57  & 0.35(5) &8.72(2)&0.86&1.08&0.46&0.36&0.60(0.45)&0.40\\ \hline
"           &4.92   & 0.50(5) &4.21(2)&0.79&0.88&0.54&0.49&0.45(0.35)&0.42    \\ \hline
"           &2.89   & 0.60(5) &2.70(3)&0.76&0.79&0.59&0.58&0.35(0.30)&0.44    \\ \hline
"           &IC(20) &        &2.06(2)&0.70&0.72&0.61&0.61&0.30(0.30)&0.42    \\ \hline
\hline
$0.06$  &23.15  & 0.27(3) &18.7(1)&0.93&1.20 &0.40&0.30&0.60(0.45)&0.35\\ \hline
"           &11.57  & 0.35(3) &9.6(1) &0.84&1.02&0.48&0.39&0.45(0.40)&0.38\\ \hline
"           &4.92   & 0.55(3) &4.35(6)&0.77&0.85&0.57&0.53&0.25(0.25)&0.40\\ \hline
"           &IC(50) &        &2.96(5)&0.68&0.71&0.59&0.57&0.25(0.25)&0.37\\ \hline
\end{tabular}
\caption[]{\label{t.res}
{\it Cooling results for IC at 20 and 50 cooling sweeps
and for RIC at various $\delta$. The lattice spacing ($a$), the 
cooling radius ($r(\delta)$), 
distances and sizes are given in fm.
The density $\langle N/V \rangle$ is in fm$^{-4}$. The errors are statistical,
when they are not explicitly quoted they amount to 1 or less
in the last indicated digit. 
}}
\end{center}
\end{table}

We are mostly interested in studying the performance of RIC with
respect to I-A pairs. 
First we observe that the density of instantons ($N/V$ in Table \ref{t.res})
increases with $\delta$. This corroborates the finding that less I-A annihilation
takes place for large $\delta$. Moreover the density scales correctly 
(up to at most $10\%$ ($20\%$ for the largest $\delta$) difference) by
comparing 
the results for the two lattice spacings at fixed $\delta$. 

\begin{figure}[tb]
\vspace{12.0cm}
\includegraphics{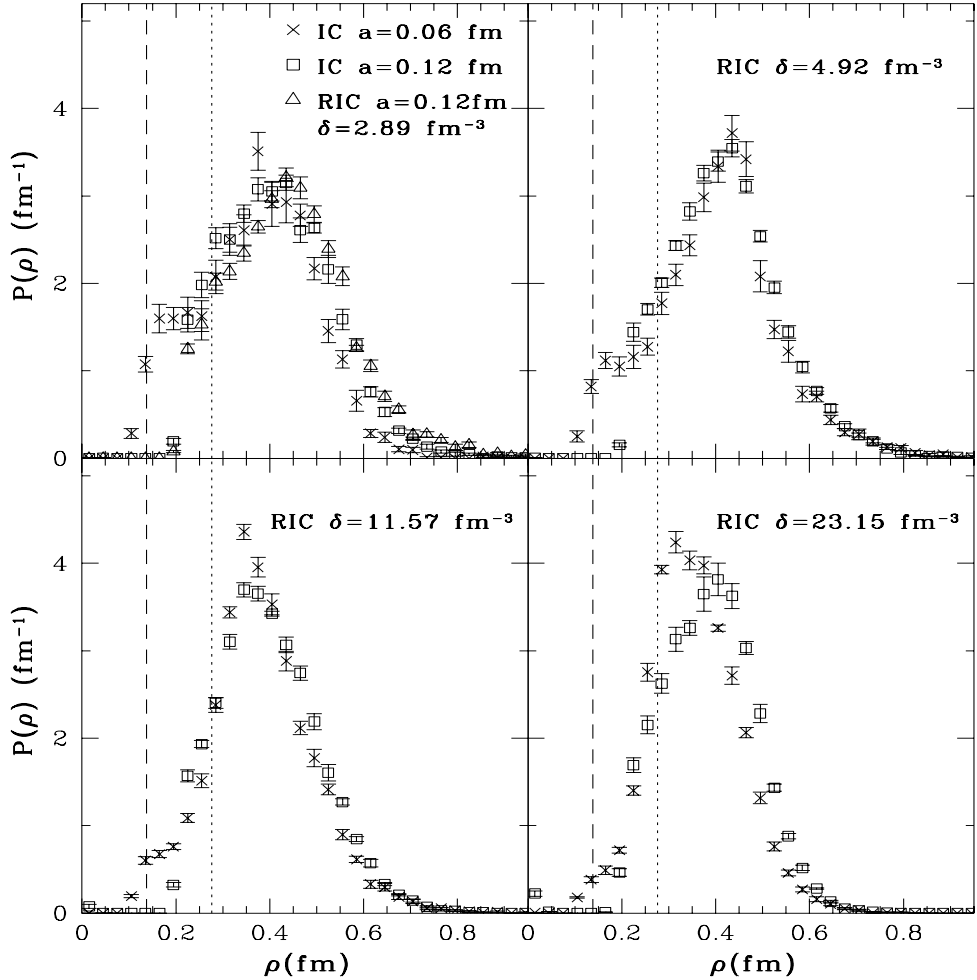}
\caption{{\it Size distributions on the $a=0.12$ fm lattice (squares, triangles)
and $a=0.06$ fm lattice (crosses).
The vertical dotted (dashed) lines indicate the location of the IC dislocation
threshold $\rho_0=2.3a$ on the $a=0.12$ $(a=0.06)$ fm lattice.
}}
\label{f.size}
\end{figure}

We define pairs in this ensemble by assigning to each 
instanton the anti-instanton that has maximal overlap
$O_{IA}$ with it; $O_{IA}$ is defined as:
\be
\label{e.over}
O_{IA}= \frac{\rho_I+\rho_A}{2 d_{IA}},
\ee
where $\rho_{I,A}$  are the sizes and $d_{IA}$ the distance between I and A.
The results for the average overlap and separation 
are given in  Table \ref{t.res}.
As expected, large $\delta$ preserves stronger overlapping pairs.
Notice the good agreement between the pair separation $d_{IA}$ and
the cooling radius $r(\delta)$, Eq. (\ref{e.cscale}), 
a clear indication that smoothing has been active in 
removing pairs with a separation
smaller than the cooling radius.
As we will further discuss below, this also indicates that there does not seem
to be an intrinsic scale for I-A separation apart from the one given by  $r(\delta)$
itself.

For comparison we also present in Table \ref{t.res} the overlap and separation
between maximally overlapping objects of the same charge.
Note that for large $\delta$, i.e. denser ensemble, 
objects with opposite charges are closer on average than 
objects with equal charges. 
However, in our earlier IC analysis \cite{mnpNP}, which roughly
corresponds to RIC at small  $\delta$,
we observed a homogeneous distribution of I and A over the lattice. The
difference can be understood by noticing that pairs with strong overlap
will be removed faster during cooling, and have probably been eliminated at
the stage of smoothing provided by IC.
Since the action of an I-A pair decreases as the overlap increases and
is smaller than that of an I-I or an A-A pair, 
it is indeed energetically favourable
that I-A pairs overlap more.  

In Fig. \ref{f.size} we present the instanton size distribution
as a function of $\delta$, Table \ref{t.res} provides the average instanton size
$\langle \rho \rangle$.
Since IC and RIC converge to scale-invariant instanton configurations,
the instanton sizes, in contrast to other physical quantities,
should not change when the cooling radius increases beyond them
(up to modifications that would be induced by I-A annihilation and possible
dressing of the instantons by fluctuations).
This is certainly the case for the results obtained with IC above the minimal 
smoothing level (see \cite{mnpNP}) and also for RIC with small $\delta$
(i.e. density of instantons below $4$ fm$^{-4}$).
Note, however, that now we go to much denser and rougher
ensembles. At these densities the stability  of the size distribution 
is not as apparent, both the average size and the shape of the
distribution weakly depend on the smoothing level.
The increase in the close pair population
seems to lead to a certain shift in the average size, both by spoiling the size
estimate and by re-weighting the size distribution\footnote{Note that the size
distributions of Fig.~\ref{f.size} are normalized, while describing
instanton populations varying by a factor up to 10.}.
It is difficult to evaluate which of these two effects dominates, cf. 
the debate about this problem in the literature
(for reviews, see \cite{Pierre, Neg}).
According to \cite{boulder} smoothing the configurations induces
a modification of the size and the correct one should be obtained by
linear extrapolation to zero smoothing levels (for SU(3) this seems to give 
rough agreement between the results from different smoothing techniques \cite{Neg}). 
Our attitude is not to trust the size determination corresponding to large densities
(small cooling radius), since, as we have discussed, it is difficult in
such cases to obtain a reliable description of the action and charge density
in terms of instantons (the deviations amount to 0.6 (0.5) of the total charge
(action)).

This problem cannot be easily solved and is indeed farther reaching,
since it is related to the general question of whether there exists
an effective I-A density relevant for a given physical effect.
The numerical agreement between the cooling radius and the I-A separation
-- see Table~\ref{t.res} -- suggests that the latter is determined by
the former, i.e. there appears to be no intrinsic scale for I-A separation
independent of the one set by the smoothing scale itself.

\begin{figure}[tbp]
\vspace{6.5cm}
\includegraphics{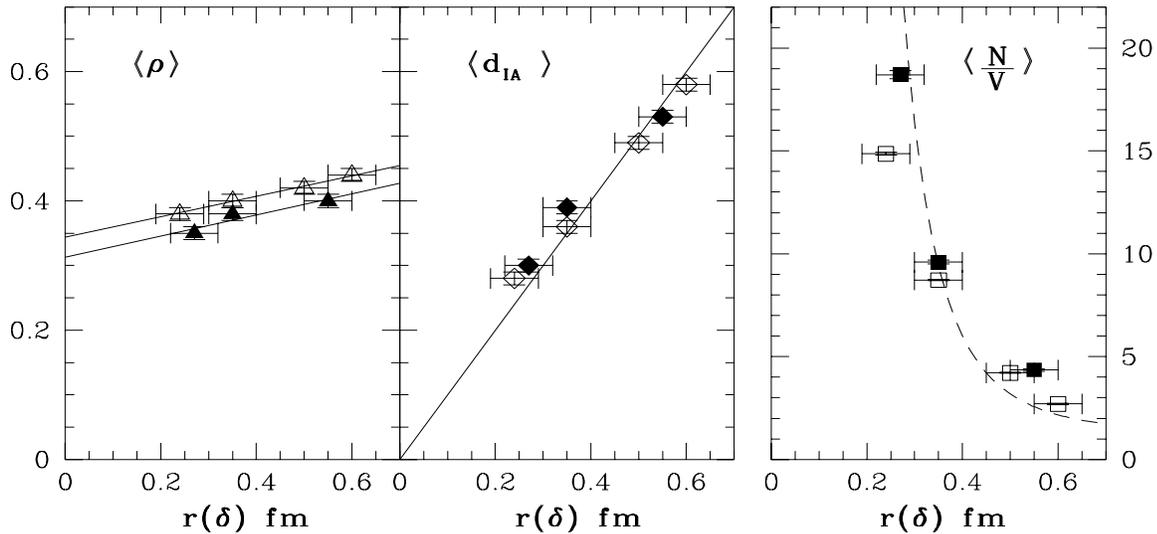}
\caption[]{\label{f.extr}
{\it Dependence of the average size (triangles),
the I - A average distance (diamonds)
and the instanton density (squares)
on the cooling radius $r$. 
The dashed curve fits the density data to 
$\langle N/V\rangle= \langle N/V\rangle_{\rm asymptotic}
+ C/r^4$.
Open (filled) symbols correspond to the $a=0.12$ $(0.06)$ fm lattice.
}}
\end{figure}

In order to elaborate on this, we consider again the instanton properties 
as a function of the cooling radius. 
It is apparent from Fig. \ref{f.size} and Fig. \ref{f.extr} that the
dependence of the average size upon $r(\delta)$ is
rather weak, even at small $r(\delta)$ (large $\delta$).
This seems to indicate the existence of a non-vanishing, 
dominant instanton size even
at cooling radius zero. However, Fig. \ref{f.extr} also shows
the different behaviour of the density 
$\langle N/V \rangle$ and of the pair separation $\langle d_{IA}\rangle$.
A simple extrapolation of the data to zero cooling radius is compatible with a
diverging density and a vanishing I-A distance. 
This is in accord with the fact that there is no principal 
distinction between a trivial 
fluctuation and an instanton--anti-instanton pair,
apart from the somewhat arbitrary criterion given in terms of $S^{\rm IA}_{\rm int}$. 
This supports the observation made above that there appears to be
no average I-A separation or instanton density that could
be defined independently of the smoothing scale (at least up to the
scales we have probed).
Thus, speaking of an instanton ensemble seems to be possible only 
in the context of smoothing and by providing a specific smoothing scale.
If this is indeed the case, a criterion for selecting ``physically relevant" pairs
may only come from considering the relevant scale of the phenomena
to which we can expect these pairs to contribute.

\section{Conclusions and Outlook}

By fixing the amount of cooling with the help of a {\it cooling radius} 
($r(\delta)$) above which physical observables such as the string tension
remain untouched, restricted improved cooling provides a {\it smoothing scale}
for treating MC 
configurations. Because of its connection with the equations of motion this
scale appears as a continuum length scale, which can be systematically 
determined from the measurement of one physical quantity. 
Starting from the rough, 
UV-dominated field configurations (cooling radius $r=0$), RIC with 
increasing $r$ acts as a ``low-pass filter" by smoothing out wavelengths
smaller than $r$ and producing configurations on which correlation
functions at distances beyond $r$ retain the physics of the larger 
wavelengths. 
The conjecture that the cooling radius $r(\delta)$
provides a universal smoothing scale seems well supported by our analysis.

The observation of topological structure itself poses a
difficult problem. The topological susceptibility appears to be
quantitatively under control, 
but the more detailed topological structure raises many questions.
Firstly, for very small cooling radius the observed structure
is very dense and only poorly described in semi-classical terms.
Secondly, we find that cooling uncovers
topological structure on scales close to the cooling radius.
While the instanton size distributions
 are rather stable under variations of the
smoothing scale, the instanton density and the mean separation 
of instanton--anti-instanton pairs appear to be determined by it.
Decreasing the smoothing scale, this structure thus becomes 
denser until it continuously disappears into the UV fluctuations.
Hence, for these quantities,
an extrapolation to no smoothing does not lead to meaningful
results.
On the other hand, it appears to be quite arbitrary 
to choose from the observed structure a ``physical"  sub-ensemble by fixing a
finite smoothing scale. 
It may well be that the characteristic length scale of the 
phenomena under scrutiny
has to be taken into account when establishing the criteria for finding an
``effective" relevant structure that contributes to these phenomena. 
Here RIC may 
provide a useful tool, since it introduces a smoothing radius as
a continuum length scale, which can
be  systematically varied in the analysis. 

Further studies should provide improved statistics for the precise 
determination of 
the cooling radius and the other properties of RIC, but also a better
theoretical understanding of these questions.

\section*{Acknowledgements}
It is a pleasure to thank Ferenc Niedermayer for suggesting the 
modification of  
cooling  that initiated this study, and for discussions.
We also thank Pierre van Baal for a critical reading
of the manuscript and discussions.
Part of this work has been performed while M.G.P. and I.O.S. 
were members of the ISKOS workshop at ZiF, Bielefeld.
Partial support from the European Network ``Finite Temperature Phase 
Transitions
in Particle Physics" and from DFG is gratefully acknowledged. 
M.G.P. would also like to acknowledge support from CICYT and 
hospitality at CERN's Theory Division while part of this work was being
performed.

\section*{Appendix}

We provide a few details about the way we perform the extraction of the instanton content from
the configurations (a more detailed description of the method will be presented
elsewhere).

As already mentioned in section 5, our instanton identifier relies on the 
following assumptions:
\begin{enumerate}
\item
Instantons are  self-dual objects with energy density localized in space and time.
Self-duality imposes, in particular, that the localization of the peaks  
should be the same 
in the electric and magnetic parts of the action density and also in the charge
density (for details on how we imposed the cut from self-duality, see
\cite{mnpNP}).
\item
We  approximate the action and charge
density by a superposition of self-dual non-interacting instantons and
anti-instantons parametrized through the analytic expression for the 1-instanton
action and charge density:
\bea
s_i(x) &=& {48 \over {  (\rho^i)^4}}
\left[1+\sum_{\mu=1}^4 \left( {{x_\mu - c_\mu^i} \over {\rho^i}}\right)^2
\right]^{-4},
\label{e.Qc} \\
s_{\rm fit}(x)&=& \sum_{i=1,N} s_i(x) \ ; \ \
q_{\rm fit}(x)= \frac{1}{8\pi^2}
\sum_{i=1,N} n_i s_i(x),
\eea
with $N=N_I+N_A$ the number of instantons plus anti-instantons, $\rho^i$
and $c_\mu^i$ respectively the size and location of the $i$th (anti)instanton
and $n_i=1$  for an instanton, $n_i=-1$ for an anti-instanton\footnote{
To account for periodicity effects we add to the energy density of
each instanton the eight closest replicas, obtained by displacing (\ref{e.Qc})
by a torus period in one of the four directions.}.
\end{enumerate}

 The departure of the real action and charge from the above formula is measured
through the quantities $\varepsilon_{q,s}$ given in Eq. (\ref{e.dfit}).
For the analysis in this paper we have supplemented the one performed in
\cite{mnpNP} with a further restriction: every time an instanton candidate
is located, it is counted only if, by adding it to $s_{\rm fit}$
and $q_{\rm fit}$, $\varepsilon_{s,q}$ simultaneously decrease.
Once the instanton has been accepted, its contribution ($s_i(x)$) is
subtracted from the total density before proceeding to analyse
the next instanton candidate.
Note that this procedure assumes that all self-dual
objects in the ensemble can be described in terms of a superposition
of instantons following the BPST ansatz (other possible
self-dual objects, which  are not well parametrized by
this ansatz, would be discarded by our $\varepsilon_{s,q}$ cuts).
A small value of $\varepsilon_{s,q}$  is, hence, a signal
that most of the structure present in the ensemble 
is well described in these terms.

As an example of the kind of results we expect, let us consider
 the I-A pairs in Fig. \ref{f.ia}, where $\varepsilon_s$ ($\varepsilon_q$)
varies from
about 0.1 (0.05), for the pair with smallest $S_{\rm int}^{\rm IA}$ in the figure,
to 0.32 (0.18), for the one with largest $S_{\rm int}^{\rm IA}$. Beyond the
latter,
our instanton finder fails to identify I and A any longer.
In the same way, for the small instantons in Fig.~\ref{f.actr},
$\varepsilon_s$ ($\varepsilon_q$)
goes from  0.05 (0.15)  for the largest one in the figure to 0.06 (0.86)
for the smallest one. Again beyond that point our finder no longer
sees an instanton at this location.


%


\begin{thebibliography}{99}
\bibitem{alb}
M.~Albanese et al.,
Phys.~Lett. B192 (1987) 163; Phys.~Lett. B197 (1987) 400.\\
M. Teper, \PLB 183 (1987) 345; \PLB 185 (1987) 121.  
%
\bibitem{digiat}
C. Christou, A. Di Giacomo, H. Panagopoulos and E. 
Vicari, \PRD 53 (1996)  2619;  B. Alles, M. D'Elia and A. Di Giacomo,
\NPB 494 (1997) 281; Nucl. Phys. Proc. Suppl. 53 (1997) 541; 
\PLB 412 (1997) 119. 
%
\bibitem{boulder}
T. De Grand, A. Hasenfratz and T. G. Kovacs, \NPB 520 (1998) 301.
A. Hasenfratz and C. Nieter, hep-lat/9806026; hep-lat/9810067.
%
\bibitem{phil}
P. de Forcrand and J. E. Hetrick, Nucl. Phys. Proc. Suppl. 63 (1998) 838.
%
\bibitem{tep0}
B. Berg, \PLB 104 (1981) 475;  Y. Iwasaki and T. Yoshie,
\PLB 131 (1983) 159; S. Itoh, Y. Iwasaki and T. Yoshie, \PLB 147
(1984) 141; M. Teper, \PLB 162 (1985) 357; 
J. Hoek, M. Teper and J. Waterhouse, \NPB 288 (1987) 889.
\bibitem{schi}
E. M. Ilgenfritz, M. L. Laursen, G. Schierholz, M. Muller-Preussker,
H. Schiller, \NPB 268 (1986) 693.
\bibitem{poli}
M. I. Polikarpov and A. I. Veselov, \NPB 297  (1988) 34.
\bibitem{michael}
C. Michael and P. S. Spencer, \PRD  52 (1995) 4691.
\bibitem{negele}
M. C. Chu, J. M. Grandy, S. Huang and J. W. Negele, Phys. Rev. D49 (1994) 6039.
\bibitem{gon95}
A. Gonz\'alez-Arroyo, P. Mart\'{\i}nez and A. Montero, Phys.~Lett. B359
(1995) 159;\\
A. Gonz\'alez-Arroyo and A. Montero, Nucl. Phys. Proc. Suppl. 47 (1996) 294;
Phys. Lett. B387 (1996) 823;
Nucl. Phys. Proc. Suppl. 53 (1997) 596.
\bibitem{mnp2}
P. de Forcrand, M. Garc\'{\i}a P\'erez and I. O. Stamatescu,
Nucl. Phys. Proc. Suppl. 47 (1996) 777; Nucl. Phys. Proc. Suppl. 53 (1997) 557. 
%
\bibitem{mnpNP}
P. de Forcrand, M. Garc\'{\i}a P\'erez and I. O. Stamatescu,
\NPB 499 (1997) 409.
%
\bibitem{mnp3}
P. de Forcrand, M. Garc\'{\i}a P\'erez, J. E. Hetrick and I. O. Stamatescu,
Nucl. Phys. Proc. Suppl. 63 (1998) 549, hep-lat/9802017.
\bibitem{ts98}
M. Teper, ``Physics from the lattice: Glueballs in QCD: Topology: SU(N)
for all N" 
in {\it Cambridge 1997, Confinement, duality, and nonperturbative 
aspects of QCD}, p. 43 (hep-lat/9711011);
D. A. Smith and M. Teper, Phys. Rev. D58 (1998) 014505.  
\bibitem{alvaro}
A. Montero, `` Confinamiento y configuraciones cl\'asicas", PhD thesis
 (1998), Universidad Aut\'onoma de Madrid.
\bibitem{boulder0}
T. DeGrand, A. Hasenfratz and D. Zhu, Nucl. Phys. B475 (1996) 321;
Nucl. Phys. B478 (1996) 349. T. DeGrand, A. Hasenfratz and T. G. Kovacs,
Nucl. Phys. B505 (1997) 417.
\bibitem{mpr}
M. Feurstein, E. M. Ilgenfritz, M. M\"uller-Preussker and S. Thurner,
Nucl. Phys. B511 (1998) 421. E. M. Ilgenfritz et al., 
Nucl. Phys. Proc. Suppl. 63 (1998) 480; 
E. M. Ilgenfritz et al., \PRD 58 (1998) 094502;
E. M. Ilgenfritz and S. Thurner, hep-lat/9810010.
\bibitem{Pierre}
P. van Baal, Nucl. Phys. Proc. Suppl. 63 (1998) 126. 
\bibitem{Neg}
J. W. Negele, hep-lat/9810053.
\bibitem{tepe1}
M. Teper, Nucl. Phys. B411 (1994) 855.
\bibitem{pisa1} A. Di Giacomo, E. Meggiolaro and H. Panagopoulos, 
Nucl. Phys. Proc. Suppl. 54A (1997) 343.
%
\bibitem{dig89}
M. Campostrini, A. Di Giacomo, M. Maggiore, H. Panagopoulos and
E. Vicari, Phys.~Lett. B225 (1989) 403.
%
\bibitem{Shuryak}
T.~Sch\"afer and E.~Shuryak, Rev.\ Mod.\
Phys.\ \underline{70} (1998) 323 and references therein. 
%
\bibitem{nied}
F. Niedermayer, private communication.
\bibitem{Valley}
A. Yung, ``Valley Method for Instanton-Induced Effects in
Quantum Field Theory", in Proc. ICTP 1991 {\it Summer School on High Energy
Physics and Cosmology}, p. 580. I. Balitsky and A. Sch\"afer, 
Nucl. Phys. B 404 (1993) 639.
\bibitem{pm}
P. J. Braam and P. van Baal, Commun. Math. Phys. 122 (1989) 267.
\bibitem{overi}
M. Garc\'{\i}a P\'erez, A. Gonz\'alez-Arroyo, J. Snippe and P. van Baal,
Nucl. Phys. B413 (1994) 535.

\bibitem{map}
M. Garc\'{\i}a P\'erez and P. van Baal, Nucl. Phys. B429 (1994) 451.
%
\bibitem{for85}
P. de Forcrand G. Schierholz, H. Schneider and M. Teper,
Phys.~Lett.~B160 (1985) 137.
\bibitem{mite} C. Michael and M. Teper, Phys. Lett. B199 (1987) 95, 
Nucl. Phys. B305 (1988) 453;
J. Fingberg et al., Nucl. Phys. B392 (1993) 493.

\bibitem{pt}
D. J. R. Pugh and M. Teper, Phys. Lett. 224B (1989) 159.

\bibitem{erha}
E. Seiler and I.-O. Stamatescu, preprint MPI-PAE/PTh 10/87.
\end{thebibliography}
\end{document}